\documentclass[twocolumn,aps,amsmath,amssymb,floatfix,prl,superscriptaddress,nofootinbib]{revtex4-2}

\usepackage{bbm}
\usepackage[dvips]{graphicx}
\usepackage{amsmath,amssymb,latexsym}
\usepackage{placeins}
\usepackage{xcolor}
\renewcommand{\ol}[1]{\overline{#1}}

\newcommand{\N}{\mathcal{N}}

\renewcommand{\Re}{\mathrm{Re}\,}
\renewcommand{\d}{\mathbf{d}}
\newcommand{\e}{\mathbf{e}}
\newcommand{\f}{\mathbf{f}}
\renewcommand{\k}{\mathbf{k}}
\renewcommand{\l}{\mathbf{m}}
\renewcommand{\r}{\mathbf{r}}

\renewcommand{\u}{\mathbf{u}}
\newcommand{\x}{\mathbf{x}}

\newcommand{\Q}{\mathbf{Q}}
\newcommand{\micron}{\mu\mathrm{m}}
\newcommand{\comma}{\quad,}
\newcommand{\period}{\quad.}
\renewcommand{\Im}{\mathrm{Im}\,}

\newcommand{\zero}{{\boldsymbol{0}}}
\newcommand{\K}{\mathbbm{K}}
\newcommand{\Kp}{\mathbbm{K}^\ast}
\newcommand{\jol}{{\overline{\jmath}}}

\renewcommand*{\vec}[1]{\mathbf{#1}}
\newcommand{\wt}[1]{\widetilde{#1}}

\newcommand*{\F}{\boldsymbol{\Phi}}

\newcommand*{\Gmat}{\boldsymbol{\Gamma}}

\newcommand*{\FP}{\boldsymbol{\Phi}^\ast}

\newcommand{\var}{\mathrm{var}}

\renewcommand{\k}{\mathbf{k}}
\newcommand{\ki}{{\mathbf{k}_{\mathrm{I}}}}

\newcommand{\kI}{{\k_\mathrm{I}}}
\newcommand{\tres}{t_{\text{res}}}

\begin{document}

\title{Synchronization in cilia carpets and the Kuramoto model with local coupling:\\
breakup of global synchronization in the presence of noise}

\author{Anton Solovev}
\affiliation{TU Dresden, 01062 Dresden, Germany \\ *\,benjamin.m.friedrich@tu-dresden.de}
\author{Benjamin M. Friedrich}
\email[ ]{ }
\affiliation{TU Dresden, 01062 Dresden, Germany \\ *\,benjamin.m.friedrich@tu-dresden.de}

\date{\today}

\begin{abstract}
Carpets of beating cilia
represent a paradigmatic example of self-organized synchronization of noisy biological oscillators,
characterized by traveling waves of cilia phase.
We present a multi-scale model of a cilia carpet,
which comprises realistic hydrodynamic interactions between cilia computed
for a chiral cilia beat pattern from unicellular \textit{Paramecium} and active noise of the cilia beat.
We demonstrate an abrupt loss of global synchronization beyond a characteristic noise strength.
We characterize stochastic transitions between synchronized and disordered dynamics, which generalizes
the notion of phase slips in pairs of coupled noisy phase oscillators.
Our theoretical work establishes a link between the two-dimensional Kuramoto model of phase oscillators with mirror-symmetric oscillator coupling
and detailed models of biological oscillators with asymmetric, chiral interactions.
\end{abstract}

\maketitle

\textbf{
Motile cilia are hair-like extensions of biological cells that beat periodically like a whip.
Collections of cilia on the surface of microorganisms or human airways
beat in a coordinated fashion as a traveling wave
-- similar to a Mexican wave in a soccer stadium -- to facilitate efficient fluid transport.
This emblematic example of synchronization of noisy biological oscillators attracted attention
from physicists and mathematicians for more than hundred years,
yet important question remain open,
including the impact of active noise on wave synchronization,
which has so far not been systematically studied for cilia carpets.
Recently, the authors proposed a multi-scale model of a cilia carpet \cite{Solovev2020b},
which in this work is advanced to include active noise of the cilia beat.
We show that global synchronization in the form of traveling waves of cilia phase
is abruptly lost at a characteristic noise strength,
and provide evidence for local synchronized patches in large systems.
We connect our findings to the celebrated Kuramoto model.
}

\subsection{Introduction}
Motile cilia and flagella are noisy biological oscillators:
the regular bending waves of these slender cell appendages exhibit
small, but measurable phase fluctuations,
causing frequency jitter
\cite{Polin2009,Ma2014,Wan2014b}.
This active noise induces phase slips in pairs of synchronized cilia,
as observed in the bi-ciliate green alga \textit{Chlamydomonas},
where one cilium occasionally executes an extra beat \cite{Goldstein2009,Goldstein2011}.
In fact, phase slips in pairs of cilia are well described by the famous Adler equation
of two coupled noisy phase oscillators \cite{Adler1946,Stratonovich1967}.

On epithelial tissues, such as in mammalian airways,
carpets of thousands of cilia can spontaneously synchronize into
so-called metachronal waves, i.e., traveling waves of cilia phase \cite{Sanderson1981}.
The coordinated beating of cilia in these cilia carpets
pumps mucus and thus clears pathogens from the airways.
Moreover, cilia carpets that exhibit metachronal synchronization are found on the surface of model organisms,
e.g., green alga colonies or unicellular \textit{Paramecium} \cite{Machemer1972,Brumley2012}.
In these systems, the traveling waves of cilia phase have a well-defined direction
relative to the common direction of the effective stroke of the beating cilia \cite{Machemer1972,Knight1954}.
In many species, including \textit{Paramecium},
the cilia beat consists of an approximately planar effective stroke during which the cilium is approximately straight,
and a recovery stroke, during which the cilium moves in a counter-clockwise fashion close to the surface,
see also Fig.~\ref{figure1}(a).
These chiral bending waves of a beating cilium reflect its internal molecular architecture,
which comprises a chiral distribution of molecular motors.
The common direction of the effective stroke in a cilia carpet is set by a global polarity of the ciliated tissue
that aligns the bases of cilia (and additionally by directed flows during tissue development) \cite{Guirao2010}.

Hydrodynamic interactions between cilia have been identified as one of the
key mechanisms for cilia coupling underlying synchronization
\cite{Taylor1952,Brumley2014}.
If exposed to moderate external flows,
the timing of the cilia beat pattern changes,
while the sequence of shapes assumed by the cilium during its beat cycle is altered only a little
\cite{Goldstein2009,Klindt2016,Pellicciotta2020}.
Importantly, if many cilia beat together, the fluid flow generated by one cilium will
either increase or decrease the hydrodynamic friction at another cilium nearby,
depending on the specific geometry of that cilium pair \cite{Vilfan2006,Solovev2020a}.
This change in hydrodynamic load subsequently changes the phase speed by which the other cilium progresses along its beat cycle and can cause synchronization
\cite{Brumley2014,Geyer2013,Klindt2016}.

A rich theoretical literature on synchronization in cilia carpets
can be broadly classified as either
detailed numerical simulations based on mechanistic models that describe the generation of cilia bending waves
\cite{Gueron1999,Elgeti2013,Stein2019,Chakrabarti2021},
or minimal models, where beating cilia are idealized, e.g., as orbiting spheres
\cite{Vilfan2006,Guirao2007,Niedermayer2008,Uchida2011,Wollin2011,Friedrich2012,Friedrich2016,Pellicciotta2020,Meng2021}.
What is missing are models of intermediate complexity
that faithfully account for the specific, chiral shape of the cilia beat,
yet still allow to average over many stochastic realizations and
systematically screen key parameters such as the noise strength of active cilia fluctuations.

The periodic sequence of cilia shapes during one beat cycle can be regarded as a limit cycle
\cite{Ma2014,Wan2014b,Werner2014}.
This limit cycle can be uniquely parametrized by a phase variable of the cilia beat
if we require that phase speed should be constant in the absence of perturbations and noise \cite{Pikovsky2003}.
Correspondingly, we can describe each beating cilium as a phase oscillator \cite{Ma2014,Wan2014b},
and a cilia carpet as an array of phase oscillators.

Here, we built on a recently established multi-scale model of a cilia carpet \cite{Solovev2020b}
to systematically study the impact of active noise in arrays of hydrodynamically coupled cilia.
The underlying modeling framework termed \textit{Lagrangian Mechanics of Active Systems}
describes beating cilia as effective phase oscillators,
where phase is used as a generalized coordinate,
whose dynamics is determined by a balance of conjugate active and passive forces \cite{Solovev2020a}.
Using this stochastic model, we address in particular the co-existence of multiple synchronized states,
corresponding to traveling wave solutions with different wave vectors,
and their prevalence in the presence of noise.
We show how the asymmetric coupling arising from a chiral cilia beat
results in a dominant wave with non-trivial wave vector as long as the noise strength remains below a characteristic noise strength.
Above this characteristic noise strength,
global synchronization is lost,
but dynamic domains of local synchronization may persist,
analogous to the two-dimensional Kuramoto model~\cite{Rouzaire2021,Sarkar2021a}.
Our work links an extensive literature on the Kuramoto model with local coupling and noise~\cite{Shinomoto1986,Doerfler2014,Sarkar2021a}
(or quenched frequency disorder~\cite{Sakaguchi1987,Strogatz1988,Acebron2005,Lee2010, Ottino2016})
to more detailed models of cilia carpets with asymmetric oscillator coupling.

 \subsection{Cilia carpet model}

In a recent publication \cite{Solovev2020b},
we had put forward a multi-scale model of a cilia carpet,
using an experimentally measured cilia beat pattern and detailed hydrodynamic computations,
albeit without accounting for active phase fluctuations of the cilia beat yet.
For the convenience of the reader,
we briefly review this multi-scale cilia carpet model
and then show how active noise of cilia beating can be naturally incorporated into this model.

\begin{figure}
\includegraphics[width=\linewidth]{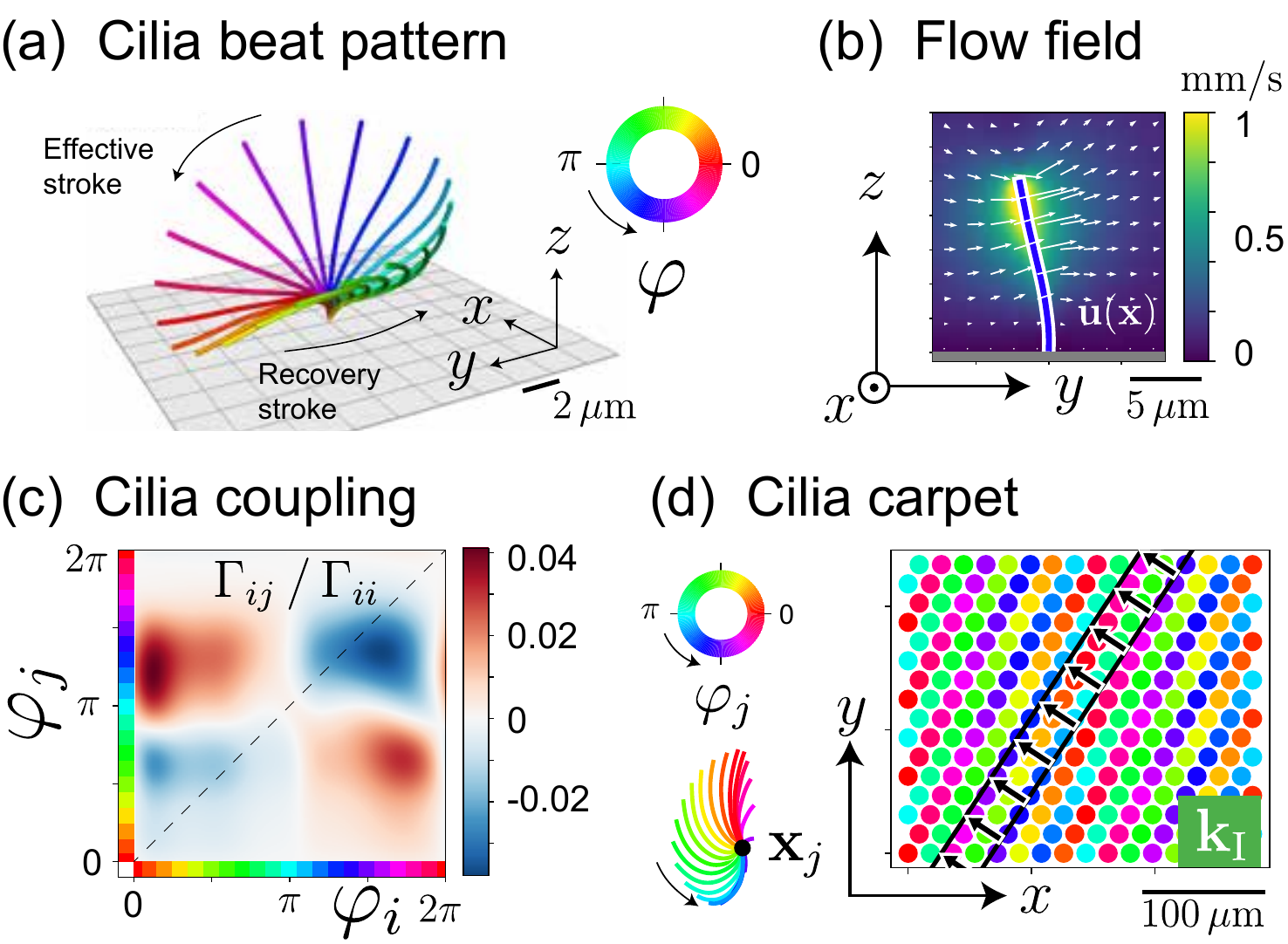} \caption{
\textbf{Multi-scale model of hydrodynamic synchronization in cilia carpets.}
(a) Cilia beat pattern from \cite{Machemer1972,Naitoh1984},
parameterized by $2\pi$-periodic phase $\varphi$.
(b) Hydrodynamic flow field $\u$ computed for this beat pattern
(color represents speed $|\u(\x)|$, arrows show projection of $\u$ on $yz$-plane; cilia phase $\varphi{=}1.4\pi$).
(c) Hydrodynamic interaction
$\Gamma_{ij}(\varphi_i, \varphi_j)/\Gamma_{ii}(\varphi_i)$
between two cilia
with separation $\x_j-\x_i = a\,(\cos\psi\,\e_x + \sin\psi\,\e_y)$, $\psi{=}\pi/3$,
as a function of their phases $\varphi_i$ and $\varphi_j$:
positive values cause cilium $i$ to beat slower.
(d) Carpet of $N$ cilia with phases $\varphi_j$
at positions $\x_j$ on a triangular lattice with periodic boundary conditions (colored dots);
the dominant traveling wave $\varphi_j = - \ki \cdot \x_j$ is indicated, see Eq.~(\ref{eq:wave}).
Lattice spacing $a=18\,\micron$,
intrinsic cilium beat frequency $\omega_0 /(2\pi) = 32 \,\mathrm{Hz}$ \cite{Machemer1972}.
Modified from~\cite{Solovev2020b}.
}
\label{figure1}
\end{figure}

We consider a carpet of $N$ cilia positioned on a regular triangular lattice of base points $\x_j$
in a rectangular domain with periodic boundary conditions,
see Fig.~\ref{figure1}(d).
Each cilium is described as a phase oscillator whose phase $\varphi_i$ advances by $2\pi$ on each cycle.
This phase variable $\varphi_i$ parameterizes a periodic sequence of three-dimensional cilia shapes,
previously measured for \textit{Paramecium}~\cite{Machemer1972,Naitoh1984}, see Fig.~\ref{figure1}(a).
We can consider $\varphi_i$ as a generalized coordinate
in the sense of Lagrangian mechanics \cite{Solovev2020a},
introduce conjugate forces, and derive an equation of motion
in terms of a force balance of generalized forces
(similar to Onsager's variational principle \cite{Doi2011,Wang2021}).
We first define a generalized hydrodynamic friction force $P_i$ for each cilium,
conjugate to its phase variable $\varphi_i$.
Specifically, this generalized friction force can be computed as the surface integral
$P_i=\int_{\mathcal{S}_i} \! d^2 \x \, \f(\x) \cdot \partial \x/\partial\varphi_i$,
where
$\mathcal{S}_i$ is the surface of cilium $i$,
$\f(\x)$ is the surface density of hydrodynamic friction forces arising from the active motion of the cilia, and
$\partial \x / \partial \varphi_i$ is the local velocity at which a point $\x$ on cilium $i$ moves
if $\varphi_i$ where to change at unit rate (\textit{principal shape change}) \cite{Solovev2020a}.
We can compute $\f(\x)$
using the Stokes equation of low Reynolds number hydrodynamics
(which is admissible because short-range interaction dominate, see \cite{Wei2019,Solovev2020a} and the discussion in \cite{Solovev2020b}),
see also Fig.~\ref{figure1}(b).
Due to the linearity of the Stokes equation,
the generalized hydrodynamic friction forces are themselves linear in the generalized velocities $\dot{\varphi}_j$
\begin{align}
\label{eq:P}
P_i (\F, \dot{\F}) &= \sum_{j=1}^N \Gamma_{ij}(\F)\,\dot{\varphi}_j \notag \\
&\approx \Gamma_{ii}(\varphi_i)\,\dot{\varphi}_i+\sum_{j\in\mathcal{N}_i} \Gamma_{ij}(\varphi_i,\varphi_j)\,\dot{\varphi}_j \comma
\end{align}
where,
 $\F = (\varphi_1,\ldots,\varphi_N)\in\mathbb{R}^N$
is the $N$-component phase vector,
$\Gamma_{ij}$ are the generalized hydrodynamic friction coefficients
which characterize \textit{hydrodynamic interactions} between cilia  $i$ and $j$ for $i \neq j$,
and $\Gamma_{ii}$ characterizes \textit{self-friction}.

Generalized hydrodynamic friction coefficients
may in general depend on the phase of every cilium $\varphi_j$,
but not on the generalized velocities $\dot{\varphi_j}$.
In the second line of the Eq.~\eqref{eq:P},
we made a pairwise interaction approximation by averaging out the (weak) dependence
of $\Gamma_{ij}$
on the non-essential variables $\varphi_k$ with $k\neq i,j$.
Likewise, we averaged out the dependence of $\Gamma_{ii}$ on $\varphi_k$ for $k\neq i$.
We additionally restricted the sum of these hydrodynamic interactions to cilia $j$ from a neighborhood $\mathcal{N}_i$ in the vicinity of cilium $i$, because hydrodynamic interactions with cilia outside of this neighborhood turned out to be weak,
see \cite{Solovev2020a} for details.

Hydrodynamic friction corresponds to a continuous dissipation of energy at a rate
$\sum_j P_j\dot{\varphi}_j$. This energy is provided by active processes within each cilium,
which we can coarse-grain into an active driving force $Q_i=Q_i(\varphi_i)$
that drives the beat of cilium $i$.
At low Reynolds numbers,
we have a \textit{force balance}
between the generalized friction force $P_i$ and
this active driving force $Q_i$
\begin{equation}
\label{eq:force_balance}
Q_i(\varphi_i) = P_i(\F,\dot{\F}) \quad,\quad i=1,\ldots,N \quad.
\end{equation}
The active driving force $Q_i(\varphi_i)$ of each cilium
is uniquely determined by a reference condition
that the phase speed of a cilium should be constant
in the absence of perturbations or noise
(i.e., $\dot{\varphi}_i = \omega_0$, while $\dot{\varphi}_k=0$ for all $k\neq i$),
which yields $Q_i(\varphi_i)=\omega_0\,\Gamma_{ii}(\varphi_i)$.
Inverting the force balance
Eq.~\eqref{eq:force_balance}
gives the \textit{equation of motion}
$\dot{\F} = \Gmat^{-1}\cdot\Q$,
where $\Q=(Q_1,\ldots,Q_N)$ is the vector of active driving forces.

This equation of motion implies a \textit{load response} of beating cilia, i.e.,
individual cilia beat slower (or faster) if external flows increase (or decrease) the hydrodynamic friction force acting on them.
The relative change in the phase speed of cilium $i$ induced by the motion of cilium $j$
is given by the ratio
$\Gamma_{ij}(\varphi_i,\varphi_j)/\Gamma_{ii}(\varphi_i)$,
which is a function of the respective phases,
see Fig.~\ref{figure1}(c).
Previously, similar models were shown to be in quantitative agreement
with the measured load response in the green alga \textit{Chlamydomonas}~\cite{Klindt2016}
as well as cilia bundles exposed to external flows~\cite{Pellicciotta2020}.

\subsection{Active noise}
To account for active phase noise of beating cilia \cite{Goldstein2009,Ma2014},
we augment the equation of motion
independent Gaussian white noise terms $\xi_j(t)$
to the equation of each $\dot{\varphi}_j$,
\begin{equation}
\label{eq:noise}
\dot{\F} = \Gmat^{-1}(\F)\cdot\Q(\F) + \Xi(t),
\end{equation}
where $\Xi(t)=(\xi_1(t),\ldots,\xi_N(t))$ is a noise vector
with independent Gaussian white noise components $\xi_i(t)$ and
$\langle \xi_i(t)\xi_j(t')\rangle = 2D\,\delta_{ij}\,\delta(t-t')$
with noise strength $D$.

\subsection{Kuramoto order parameter}
We consider perfect traveling waves
that are compatible with the periodic boundary conditions of the cilia carpet model
\begin{equation}
\label{eq:wave}
\F_\k(t): \varphi_j(t) = \omega_\k\, t - \k \cdot \x_j \quad;
\end{equation}
here, $\vec{k}$ is a wave vector corresponding to one of the $N$ reciprocal lattice points in the Brillouin zone of the cilia lattice~\cite{Solovev2020b}.

We introduce a generalized Kuramoto order parameter
by generalizing a previous definition for oscillator chains~\cite[Eq.~(160)]{Gupta2014}
\begin{equation}
\label{eq:rk}
r_\k(\F) =
N^{-1}
\left|
{\textstyle \sum_j}
\exp i(\varphi_j + \mathbf{k}\cdot\mathbf{x}_j )
\right|
\quad.
\end{equation}
Note that
$r_\k(\F_\k)=1$,
but
$r_\k(\F_\l)=0$ for $\l\neq\k$.

In the absence of noise with $D=0$,
the cilia carpet model exhibits traveling wave solutions
$\F_\k^\ast$
for each admissible wave vector $\k$,
which are not exactly equal to the perfect waves $\F_\k$, but close \cite{Solovev2020b}.
For these so-called \textit{metachronal wave solutions}, we have
$r_\k(\F_\k^\ast)\approx 1$.
Several wave solutions are linearly stable to small perturbations.
Below, Fig.~\ref{figure3}(a) shows for each stable wave solution $\FP_\k$
the maximal real part of the dimensionless Lyapunov exponents,
corresponding to the slowest decaying perturbation mode
(with relaxation time $\tau \approx -T_0 / \max \Re\lambda_j$, where $T_0=2\pi/\omega_0$ is the intrinsic period of the cilia beat).
Despite this multi-stability of wave solutions,
the majority of trajectories with random initial conditions converges to a dominant wave $\ki$ \cite{Solovev2020b}, shown in Fig.~\ref{figure1}(d).

\begin{figure*}
\includegraphics[width=170mm]{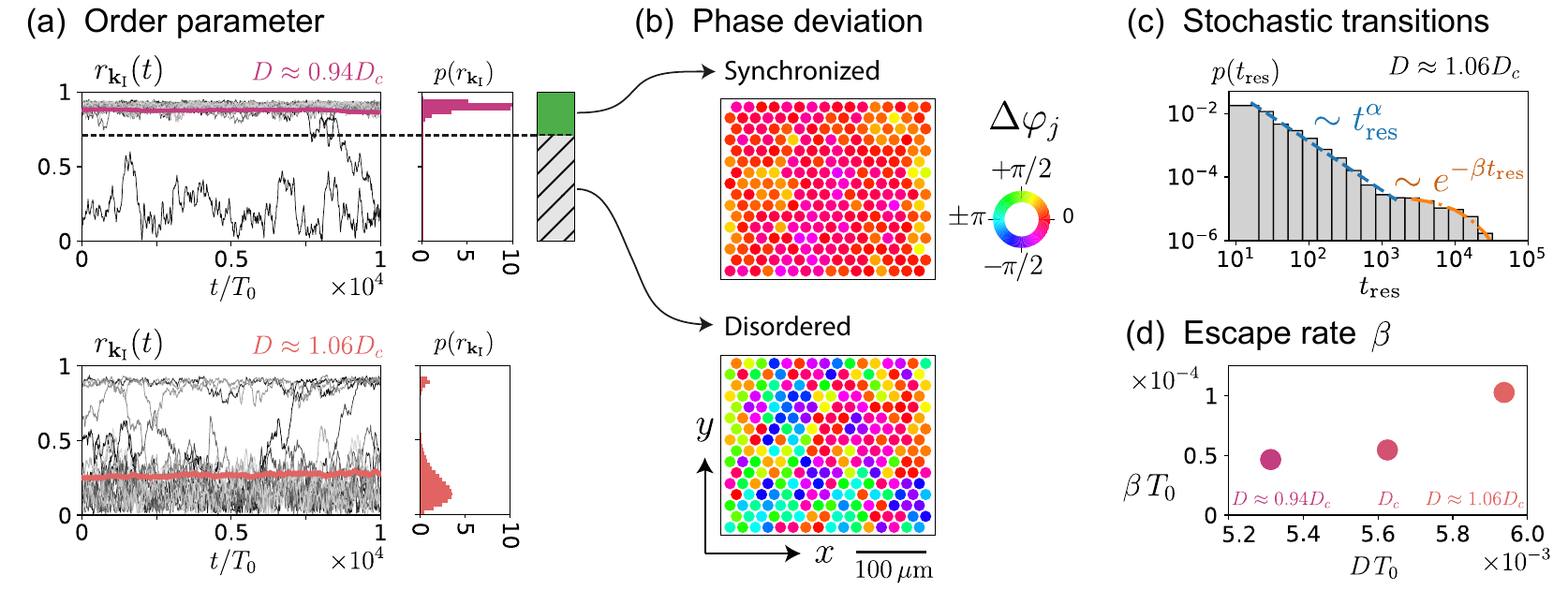}
\caption[]{
\textbf{Noise-induced transitions transitions between synchronized and disordered dynamics.}
(a)
Stochastic trajectories of the
Kuramoto order parameter $r_{\ki}(t)$
as defined in Eq.~(\ref{eq:rk}) for the cilia carpet model at steady state
for the dominant wave $\ki$ (gray colors, $n=20$)
with ensemble mean $\overline{r}_{\ki}$ (color, mean$\pm$s.e.m., $n=200$)
for two noise strengths $D \approx 0.94 D_c$ and $D \approx 1.06 D_c$.
To the right, histograms of $r_{\ki}$ are shown.
We define a \textit{synchronized} and a \textit{disordered} regime, respectively,
using a suitable threshold $r_{\ki}\gtrless 2^{-1/2}$ (dashed line).
Characteristic noise strength $D_c = 0.18\,\mathrm{s}^{-1}$.
(b)
Typical examples for
the \textit{synchronized} regime (\textit{top}, $r_{\ki} \approx 0.91$), and 
the \textit{disordered} regime (\textit{bottom}, $r_{\ki} \approx 0.19$)
depicted in terms of the deviations $\Delta \varphi_j$
of cilia phases from the dominant wave $\ki$.
Dot positions correspond to cilia positions
$\x_j$ in the $xy$-plane, $\Delta \varphi_j = \varphi_j + \ki \cdot \x_j - \langle\varphi_i + \ki \cdot \x_i\rangle$ shown in color code,
where the circular mean $\langle\cdot\rangle$ has been subtracted for visualization.
(c)
The distribution of residence times of stochastic trajectories
in the \textit{synchronized} regime
near the dominant wave $\ki$,
defined by the condition $\r_\ki>2^{-1/2}$.
While short residence times $\lesssim 10^3\,T_0$
follow a  power-law distribution with exponent $\alpha \approx -1.75$ (blue fit),
corresponding to brief excursions into the synchronized regime,
long residence times follow an exponential distribution (orange fit),
which defines an asymptotic escape rate $\beta$ from the synchronized regime.
(d)
The escape rate $\beta$
        increases as a function of noise strength $D$ close to the characteristic noise strength $D_c$.
For smaller or larger values of $D$, escapes are very rare and no sufficient statistics could be obtained.
Intrinsic period of cilia beat $T_0 = 2\pi / \omega_0 = 31.25\,\mathrm{ms}$~\cite{Machemer1972}.}
\label{figure2}
\end{figure*}

In the presence of noise with $D>0$,
the Kuramoto order parameter $\r_\ki$ for the dominant wave $\ki$ fluctuates.
Fig.~\ref{figure2}(a) shows the stochastic dynamics of $\r_\ki(t)$ for individual trajectories
for two values of the noise strength $D$ just below and above a characteristic noise strength
$D_c = 0.18\,\mathrm{s}^{-1}$, respectively.
(The ensemble had started from random initial conditions and had been equilibrated until
the ensemble-mean $\overline{r}_{\ki}$ converged.)
We observe a qualitative change of the steady-state distribution
$p(r_\ki)$ of the order parameter at this characteristic noise strength,
with most values of $r_\ki$ close to $1$ for $D<D_c$,
but much smaller values for $D > D_c$
(with $r_\ki\sim N^{-1/2}$ for $D\rightarrow\infty$).
Accordingly, we define
a \textit{synchronized} regime by
the inequality $r_\ki(\F) > r^\ast$,
and a \textit{disordered} regime by $r_\ki(\F) < r^\ast$,
where we chose the threshold $r^* = 2^{-1/2}$
(qualitative results are independent of the choice of $r^\ast$).
We observe stochastic transitions between these two regimes.
Fig.~\ref{figure2}(b) shows a typical example of
the synchronized and the disordered regime, respectively,  
in terms of the deviation of cilia phases from the perfect wave $\F_\ki$.

We computed residence times $\tres$ of individual trajectories in the synchronized regime,
i.e., the time intervals between a consecutive entrance and escape of the synchronized regime~\cite{Gammaitoni1998}.
The residence times $\tres$ follow a broad distribution:
we distinguish a regime of brief excursions into the synchronized regime
where residence times follow a power-law distribution (blue; fitted exponent $\approx{-}1.75$),
and a second regime of long residence times,
which are distributed according to an exponential distribution
$\sim\exp(-\beta\, t_\mathrm{res})$ (orange),
as expected from escape-rate theory, see Fig.~\ref{figure2}(c).
The effective escape rate $\beta$ strongly depends on noise strength $D$,
see Fig.~\ref{figure2}(d).
Note that we could not explore $\beta(D)$ for a wider range of $D$
because transitions became very rare otherwise:
for strong noise,
trajectories
rarely enter the synchronized regime,
while for weak noise,
they rarely escape.

\subsection{Noise-induced breakdown of global synchronization}

\begin{figure*}[t]
\includegraphics[width=170mm]{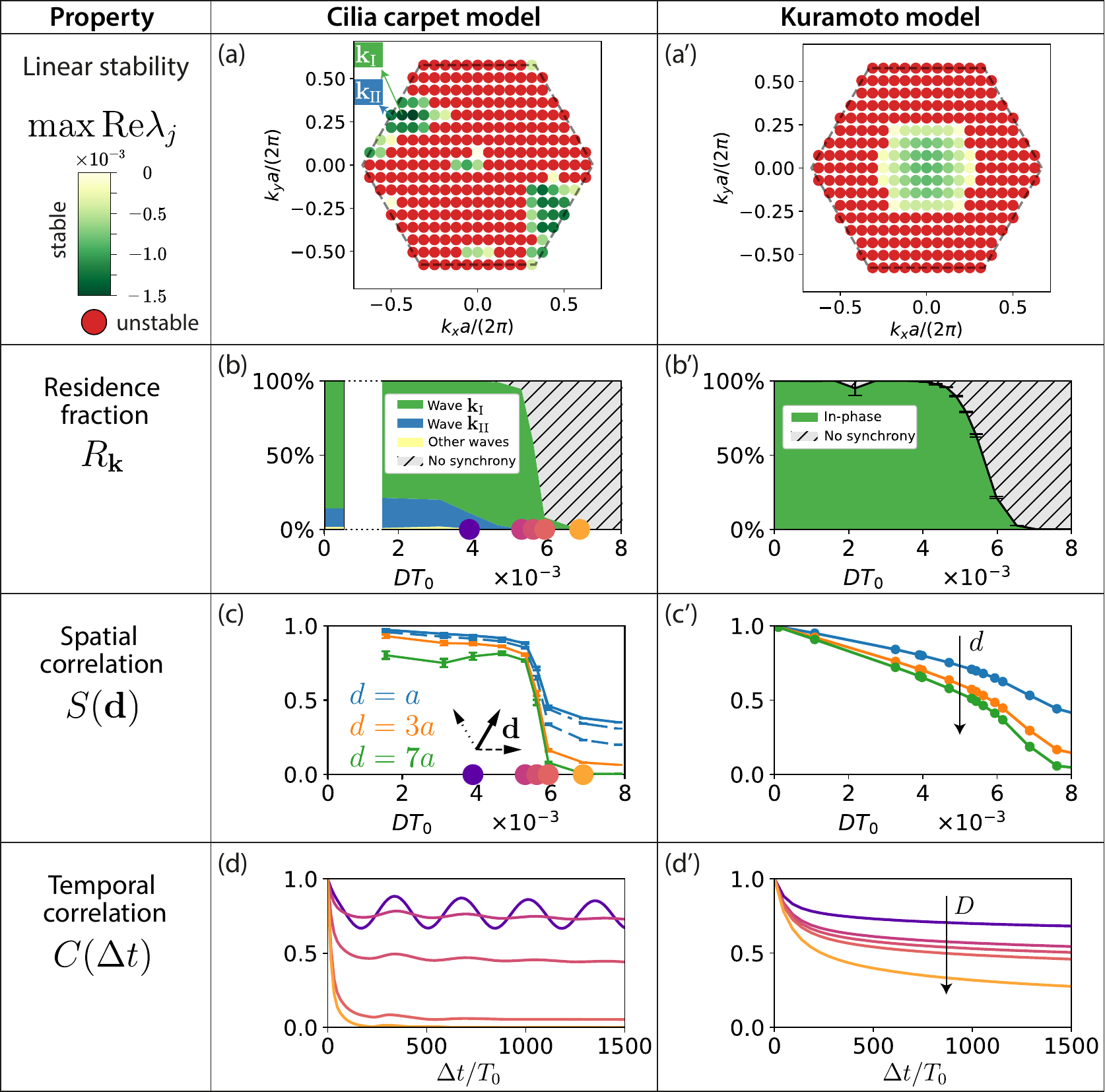} \caption{
\textbf{Noise-induced breakdown of global synchronization:}
Beyond a characteristic noise strength, global synchronization in a finite system is lost both
for the cilia carpet model (\textit{left column}) and 
the two-dimensional Kuramoto model (\textit{right column}), 
reflected by different properties (\textit{rows}).
(a)
Linear stability of traveling wave solutions
for the multi-scale cilia carpet model with asymmetric hydrodynamic interactions, Eq.~\eqref{eq:noise}.
Each colored dot represents a possible traveling wave solution
of synchronized phase dynamics with different wave vector $\k$ in the absence of noise \cite{Solovev2020b}.
The color code represents the maximal real part of the Lyapunov exponents of stable wave solutions,
which sets an inverse relaxation time of perturbations of the synchronized state.
The set of admissible wave vectors $\k$ define a Brillouin zone
that is compatible with the hexagonal lattice and the periodic boundary conditions of the rectangular unit cell
(wave vectors on opposite sides of Brillouin zone are equivalent as indicated).
(b)
Synchronized fractions: fractions of time that trajectories spend on average
in the synchronized regime for one of the waves $\FP_\k$, as a function of noise strength $D$.
The case $D=0$ corresponds to the noise-free dynamics and reflects the relative size of the basins-of-attractions of the different synchronized states.
We observe a breakdown of synchronization at a characteristic noise strength $D_c$.
Colored dots on horizontal axis indicate noise strengths considered in Fig.~\ref{figure3}(d).
(c)
Spatial correlation $S(\d)$, Eq.~(\ref{eq:two_point_correlation}),
as a function of noise strength $D$
for different distances $d=|\d|$ (colors)
and different directions (line styles, see inset).
(d)
The temporal correlation function
$C(\Delta t)$ of individual oscillators, Eq.~\eqref{eq:two_time_correlation},
for different noise strengths $D$ (with color code as in panel (b)).
Each temporal correlation function displays two regimes of
\textit{fast} decay
(due to rapid fluctuations of individual phases around a global wave)
followed by \textit{slow} decay
(due to fluctuations of global wave speed).
The apparent oscillations for weak noise in the case of the cilia carpet model reflect noise-induced oscillations around the stable spiral $\FP_\k$.
$T_0 = 2\pi / \omega_0$.
(a')-(d') Same as panels (a)-(d), but for a two-dimensional Kuramoto model,
Eq.~\eqref{eq:kuramoto}.
Coupling strength $K = 0.72\,\mathrm{s}^{-1}$ in the Kuramoto model
was chosen to match $D_c$.
}
\label{figure3}
\end{figure*}

Fig.~\ref{figure3}(a)
shows the linear stability of all possible wave solutions  $\FP_{\k}$
 in the absence of noise~\cite{Solovev2020b}.
For each wave,
we determined the \textit{synchronized fraction} for this wave in the presence of noise,
i.e., the time fraction that trajectories spend in the synchronized regime for this wave
defined by mutually exclusive conditions $r_\k>2^{-1/2}$,
see Fig.~\ref{figure3}(b).
For weak noise, results closely resemble the relative sizes of basins-of-attraction
previously determined in the absence of noise \cite{Solovev2020b}.
In particular, most trajectories spend most of their time near the dominant wave $\ki$,
consistent with $\ki$'s large basin-of-attraction.
Near the characteristic noise strength $D_c$ identified in Fig.~\ref{figure2},
the synchronized fractions abruptly drop to almost zero for all $\k$.
Concomitantly, we observe a decline of spatial and temporal correlations.
Fig.~\ref{figure3}(c) shows the \textit{spatial correlation function} $S(\d)$ of cilia dynamics
(similar to a structure factor in condensed matter physics \cite{Chaikin1995})
for different noise strengths $D$, where
\begin{equation}
S(\d) =
\left|
 \langle
    \exp{i [\varphi(\mathbf{x}_j + \mathbf{d}, t) - \varphi(\mathbf{x}_j, t)]}
 \rangle\right|
\quad.
\label{eq:two_point_correlation}
\end{equation}
Note that for a perfect traveling wave $\F_\k$ from Eq.~(\ref{eq:wave}),
we would have $S(\d)\equiv 1$ for all distance vectors $\d$.
For weak noise,
$S(\d)$ slowly decays as a function of cilia distance $d=|\d|$,
yet with a decay length that exceeds system length, corresponding to global synchronization.
Above the characteristic noise strength $D_c \approx 0.18\,\mathrm{s}^{-1}$,
$S(\d)$ is strongly reduced,
reflecting a loss of global synchronization.
Among the three principal lattice directions,
spatial correlations are slightly higher along the direction approximately parallel to the wave front of the dominant wave.

Fig.~\ref{figure3}(d) shows the \textit{temporal correlation function}
\begin{equation}
C(\Delta t)=\left| \langle \exp i [\varphi_j(t+\Delta t)-\varphi_j(t)]\rangle \right|
\period
\label{eq:two_time_correlation}
\end{equation}
For an isolated cilium with equation of motion
$\dot{\varphi}_j = \omega_0 + \xi_j(t)$,
we would have $C(\Delta t) = \exp(-D\,t)$ \cite{Ma2014}.
For a cilium in a cilia carpet,
$C(\Delta t)$ displays instead an initial fast decay,
reflecting fluctuations of $\varphi_j$ of finite amplitude around the reference wave
$\FP_\vec{k_{\mathrm{I}}}(t)$,
followed by a slow decay,
reflecting phase diffusion of the global phase with effective diffusion coefficient $D/N$
(for small $D$, see SM text).
For weak noise, $C(\Delta t)$ displays damped oscillations,
which are consistent with predictions from linear stability analysis:
the slowest decaying perturbation of $\F_\ki^\ast$
has a Lyapunov exponent with non-zero imaginary part $\Im\lambda\neq 0$,
which implies that perturbations will decay in a spiral-like fashion
with angular frequency approximately $|\Im\lambda| / T_0$.
This frequency indeed matches
the frequency of the apparent oscillations of $C(\Delta t)$,
representing an example of noise-induced oscillations.

\subsection{Synchronization in larger systems}

We simulated the cilia carpet model for a larger carpet of $48\times 16$ cilia,
which marks the largest system size currently accessible.
Note that for this large carpet,
the equilibration time to reach a steady state becomes very long ${\sim}10^5\, T_0$.
This renders simulations of even larger systems unfeasible
(and may even have implications for real cilia carpets,
where perturbations may distort the system before steady states are reached).

\begin{figure*}[t]
\includegraphics[width=170mm]{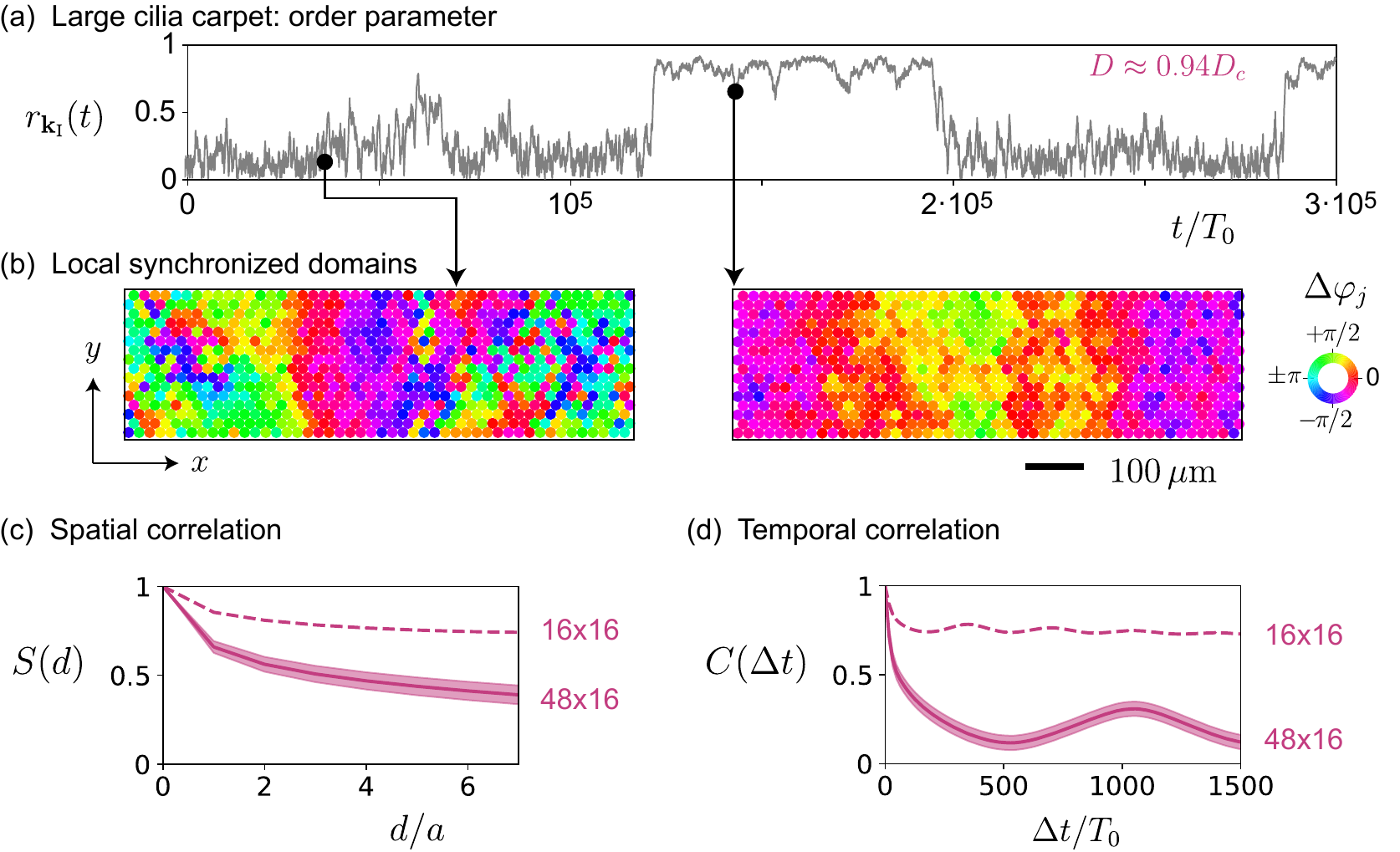}
\caption[]{
\textbf{Local synchronization in larger systems.}
(a) Kuramoto order parameter $r_\ki$ of the dominant wave $\ki$ as a function of time $t$
for a large carpet of $48\times 16$ cilia,
featuring typical transitions between synchronized and disordered dynamics.
(b) Deviations $\Delta\varphi_j$ of cilia phases from the dominant wave $\ki$
for two typical time points as indicated
($\Delta\varphi_j$ shown in color code as in Fig.~\ref{figure2}(b),
dot positions correspond to cilia positions $\x_j$ in $xy$-plane).
We find local synchronized domains,
even for phase vectors with $r_\ki<2^{-1/2}$
that would not be classified as (globally) synchronized anymore.
(c)
Spatial correlation function $S(\d)$
for a large cilia carpet of $48\times 16$ cilia (solid, mean$\pm$s.e., $\d=d\,\e_x$)
compared to that of a smaller carpet of $16\times 16$ cilia
(dashed, same as Fig.~\ref{figure3}).
(d)
Temporal correlation function $C(\Delta t)$
for large and smaller cilia carpet [analogous to panel (c)].
In larger systems, additional long-wavelength perturbation modes
cause faster decay of spatial and temporal correlations.
Noise strength $D = 0.17\,\mathrm{s}^{-1} \approx 0.94\,D_c $. }
\label{figure4}
\end{figure*}

For this $48\times 16$ lattice of cilia,
we observe again typical transitions between synchronized and disordered dynamics,
see Fig.~\ref{figure4}(a).
The deviations $\Delta\varphi_j$ of cilia phases from the dominant wave $\ki$
shown in Fig.~\ref{figure4}(b) reveal local synchronized domains,
even for low values of the Kuramoto order parameter $r_\ki$,
for which phase vectors would not be classified as globally synchronized anymore.
We observe a faster decay of the spatial and temporal correlation functions,
$S(d)$ and $C(\Delta t)$, in larger cilia carpets,
shown in Fig.~\ref{figure4}(c) and (d), respectively.
This shows that system size affects also \textit{local} properties
and is consistent with the intuitive picture that
additional long-wavelength perturbation modes with long relaxation times become excited in larger systems.
This suggests that the characteristic noise strength $D_c$ should decrease with increasing system length.
We expect that for even larger systems and intermediate noise strengths $D$
with $D_c(L_1)\ll D\ll D_c(L_2)$ (not yet accessible computationally),
cilia carpets display local synchronization on length scales $L_2$,
but not anymore on length scales $L_1$.
This statement can be made more precise by a comparison with the two-dimensional Kuramoto model.

\subsection{Comparison to Kuramoto model}
As a reference, we recall the classical Kuramoto model with local coupling~\cite{Sakaguchi1987,Doerfler2014,Sarkar2021a}
\begin{align}
\label{eq:kuramoto}
\dot{\varphi_i} = \omega_0 + \frac{K}{6} \sum_{|\x_j-\x_i|=a} \sin\left(\varphi_j-\varphi_i\right) + \xi_i(t) \quad,
\end{align}
with sinusoidal coupling between nearest neighbors on a hexagonal lattice
and independent Gaussian white noise $\xi_i(t)$ with
$\langle \xi_i(t)\xi_j(t')\rangle = 2D\,\delta_{ij}\,\delta(t-t')$.
In the absence of noise, each perfect traveling wave defined in Eq.~(\ref{eq:wave}) is a periodic solution of Eq.~(\ref{eq:kuramoto}).
A linear stability analysis reveals that multiple wave solutions are linearly stable, see Fig.~\ref{figure3}(a').
Results are highly symmetric, in contrast to the cilia carpet model.
For the in-phase synchronized state with $\k=\zero$,
the maximal real part of Lyapunov exponents $\lambda_j$ is most negative;
thus, for $\k=\zero$, the slowest decaying perturbation mode relaxes faster than for all other waves $\k$.

For the Kuramoto model, Eq.~(\ref{eq:kuramoto}),
the in-phase synchronized state $\k=\zero$ turns out to be also the dominant wave
that has the largest basin-of-attraction, see Fig.~\ref{figure3}(b') for $D=0$.
In contrast, for the cilia carpet model,
the wave for which the maximal real part of its Lyapunov exponents becomes most negative,
is only close to the dominant wave $\ki$, but slightly different \cite{Solovev2020b}.
Here, the relative size of the basin-of-attraction was determined by computing the fraction of trajectories
starting from random initial conditions (uniformly distributed) that converged to a particular wave \cite{Solovev2020b}.

We computed
the synchronized fraction for each perfect wave $\F_{\k}$
also for the Kuramoto model, Eq.~(\ref{eq:kuramoto}), see Fig.~\ref{figure3}(b').
For weak noise, virtually all trajectories remained close to the dominant wave $\k=\zero$.
For stronger noise, we observe once more a breakdown of global synchronization at a characteristic noise strength $D_c$.
The spatial and temporal correlation functions $S(\d)$ and $C(\Delta t)$
seem to decline more gradually as a function of noise strength $D$
for the Kuramoto model, Eq.~(\ref{eq:kuramoto}),
as compared to the cilia carpet model, Eq.~(\ref{eq:noise}).
Approximate analytical expression can be found for both the spatial and the temporal correlation function
by linearizing Eq.~(\ref{eq:kuramoto}), which are valid in a limit of weak noise \cite{Pikovsky2003}, see SM text.
In particular, the analytical expression for $C(\Delta t)$ is not a simple exponential decay,
but instead a product of a rapidly decaying term,
which reflects small-amplitude fluctuations of individual oscillators around the global in-phase synchronized state,
and a factor $\exp(-D\,t/N)$,
which reflects slow phase diffusion of a global phase with effective phase diffusion coefficient $D/N$,
where $N$ is the number of oscillators in the unit cell.
In short: Coupled oscillators are less noisy than a single isolated oscillator.
Finally, we do not observe any oscillatory component in the temporal correlation function for the Kuramoto model,
which is expected because all Lyapunov exponents are real in this case.

\subsection{Generalized Kuramoto models}
The most striking difference between the cilia carpet model and the Kuramoto model with local sinusoidal coupling is the fact that the dominant synchronized state is
a traveling wave $\F_\kI^\ast\approx \F_\kI$ with nontrivial wave vector $\kI\neq\zero$ for the cilia carpet,
in contrast to in-phase synchronization in the Kuramoto model.
This is a consequence of the chiral cilia beat pattern, which results in asymmetric hydrodynamic interactions.
Intriguingly, we can introduce a modified Kuramoto model with asymmetric coupling
\begin{align}
\label{eq:kuramoto2}
\dot{\varphi_i} = \omega_0 + \frac{K}{6} \sum_{|\x_j-\x_i|=a}
		\sin\left[\varphi_j-\varphi_i-\k\cdot(\x_j-\x_i)\right] \quad.
\end{align}
Note that the coupling in Eq.~\eqref{eq:kuramoto2} is
not mirror-symmetric in the sense that the coupling
acts differently on mirror-symmetric neighbors.
The variable transformation $\varphi_j \mapsto \varphi_j-\k\cdot\x_j$
maps solutions of Eq.~\eqref{eq:kuramoto} to solutions of Eq.~\eqref{eq:kuramoto2}:
this maps not only $\F_\l(t)$ to $\F_{\l+\k}(t)$,
but also all trajectories in the vicinity of $\F_\l(t)$ to
trajectories in the vicinity of $\F_{\l+\k}(t)$.
Hence, the linear stability and the basins-of-attractions
of the wave solution $\F_\l(t)$ of Eq.~\eqref{eq:kuramoto}
correspond exactly to the linear stability and the basins-of-attraction
of the wave solution $\F_{\l+\k}(t)$ of Eq.~\eqref{eq:kuramoto2}.
This implies that Fig.~\ref{figure3} applies with minor modifications
also to the modified Kuramoto model, Eq.~\eqref{eq:kuramoto2}: 
in Fig.~\ref{figure3}(a'), the green ``stability region'' would 
not be centered at the in-phase synchronized state with $\k=\mathbf{0}$ anymore 
but be shifted such that its center coincides with $\k$; 
in Fig.~\ref{figure3}(b'), the dominant synchronized fraction would not change but now correspond to wave $\k$; 
the spatial and temporal correlations shown in Fig.~\ref{figure3}(c') and (d') would not change, 
because these correlation functions quantify order irrespective of the dominant wave.

More generally, we can consider
\begin{align}
\label{eq:kuramoto3}
\dot{\varphi_i}
&= \omega_0 + \sum_{j\in\N_i} C_{ij}(\varphi_i,\varphi_j) \quad,
\end{align}
with generic coupling functions $C_{ij}(\varphi_i,\varphi_j)$
and a set of of neighbors $\N_i$ for each oscillator $i$.
We introduce the spatial mirror operation
$j\mapsto\jol$ such that $\x_j-\x_i = -(\x_\jol-\x_i)$.
In Eq.~\eqref{eq:kuramoto3}, we assume identical oscillators and translation invariance,
which implies
$C_{ij}(\varphi_i,\varphi_j) = C_{\jol i}(\varphi_i,\varphi_j)$.
A special case of Eq.~\eqref{eq:kuramoto3}
would be a \textit{phase-invariant} coupling, i.e.,
coupling functions that are invariant under a global shift of phase $\varphi_i \mapsto \varphi_i+\delta$.
In this case, Eq.~(\ref{eq:kuramoto3}) has perfect traveling wave solutions $\F_\k$.
    If the coupling is not phase-invariant, but weak,
one can exploit a separation of time-scales and
write $\varphi_j = \ol{\varphi} + \Delta\varphi_j$
as a sum of a fast variable, the global phase $\ol{\varphi}$, and
slow variables, the phase deviations $\Delta\varphi_j$ \cite{Pikovsky2003,Meng2021}.
This allows to average the coupling functions over one cycle of $\ol{\varphi}$,
which yields a phase-invariant effective coupling and
shows that we can expect to find approximate traveling wave solutions $\F_\k^\ast$ of Eq.~\eqref{eq:kuramoto3}
close to the perfect waves $\F_\k$
if the coupling is weak.
This is indeed what we find for the cilia carpet model.

\begin{figure}
\includegraphics[width=\linewidth]{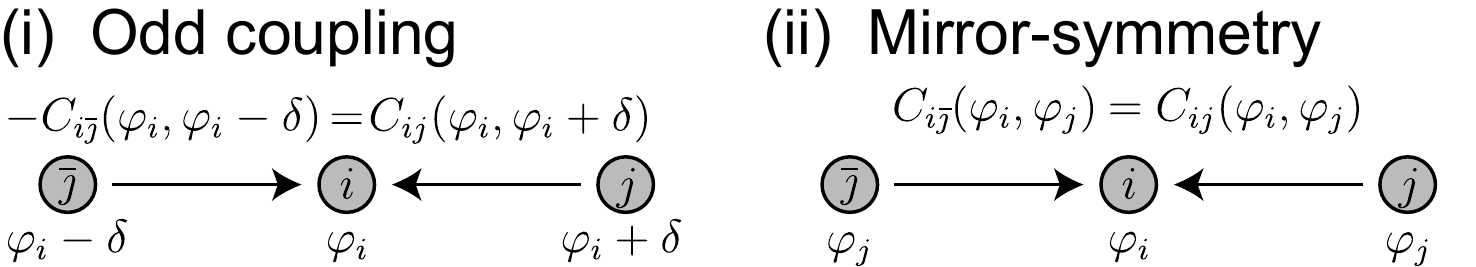}
\caption{
\textbf{Illustration of conditions (i) and (ii) for the generalized Kuramoto model.}
\textit{Odd coupling} condition (i) for the coupling functions $C_{ij}(\varphi_i,\varphi_j)$
of a generalized Kuramoto model, Eq.~\ref{eq:kuramoto3},
implies that the dispersion relation for wave solutions is trivial,
whereas \textit{mirror-symmetry} condition (ii)
implies that waves $\k$ and $-\k$ have the same stability properties.
}
\label{figure5}
\end{figure}

Two properties of the generic model of coupled oscillators given by Eq.~\eqref{eq:kuramoto3}
allow us to highlight differences between the cilia carpet model and the classical Kuramoto model, see Fig.~\ref{figure5}
\begin{itemize}
\item[(i)]
\textit{odd coupling:}
$C_{ij}(\varphi_i,\varphi_i+\delta) = -C_{i\jol}(\varphi_i,\varphi_i-\delta)$
for all $i$, $j$, and phase shifts $\delta$;
this condition implies that the traveling wave solutions $\F_\k$ of Eq.~\eqref{eq:kuramoto3} have collective frequency
$\omega_\k = \omega_0$, i.e., the dispersion relation is trivial.
\item[(ii)]
\textit{mirror-symmetry:}
$C_{ij}(\varphi_i,\varphi_j) = C_{i\jol}(\varphi_i,\varphi_j)$
for all $i$, $j$;
with this condition, we can map any trajectory $\F(t)$ of Eq.~\eqref{eq:kuramoto3}
to a new trajectory $\F'(t)$ with $\varphi_j'(t)=\varphi_\jol(t)$
defined as the mirror image of $\F(t)$.
In particular, the linear stability and basins-of-attractions of traveling wave solutions $\F_\k^\ast$
must be symmetric under the mirror operation $\k \mapsto -\k$ for a mirror-symmetric coupling.
\end{itemize}
Condition (i) may appear unusual,
but in fact reduces to the common notion of an \textit{odd} function
$\ol{C}_{ij}(\varphi_j-\varphi_i) = C_{ij}(\varphi_i,\varphi_j)$
if the coupling is phase-invariant and the mirror-symmetry condition (ii) holds.
Eq.~\eqref{eq:kuramoto2} can be considered as a minimal example that still satisfies condition (i),
but breaks the mirror-symmetry condition (ii), which is a pre-condition that a non-trivial traveling becomes the dominant traveling wave solution.

In the cilia carpet model, Eq.~\eqref{eq:force_balance},
the general coupling determined by the hydrodynamic interactions between cilia breaks both condition (i) and (ii);
correspondingly, we observe a dominant traveling wave solution, as well as a non-trivial dispersion relation
\cite{Solovev2020b}.

A comment on symmetries:
the matrix of hydrodynamic interaction coefficients
$\Gamma_{ij}(\F)$ is symmetric due to Onsager reciprocity \cite{Solovev2020a},
thus $\Gamma_{ij}(\varphi_i,\varphi_j) = \Gamma_{ji}(\varphi_j,\varphi_i)$.
But the individual coefficients are not symmetric in their arguments,
$\Gamma_{ij}(\varphi_i,\varphi_j) \neq \Gamma_{ij}(\varphi_j,\varphi_i)$.
Thus, Onsager reciprocity does \textit{not} imply the mirror-symmetry condition (ii).
In conclusion,
metachronal coordination of cilia with non-trivial dominant wave vector $\ki\neq\zero$
is only possible because the individual hydrodynamic interaction coefficients are not symmetric in their arguments.

The odd part of phase-averaged coupling functions
determine the linear stability of the in-phase synchronized state of Eq.~\eqref{eq:kuramoto3},
while the even part determines its collective frequency.
 This picture generalizes to traveling wave solutions
if one considers instead odd and even kernels of a Fourier-transformed coupling function~\cite{Meng2021}.
In this minimal model of a cilia carpet in terms of hydrodynamically coupled rotors,
the stability criterion is invariant under $\k\rightarrow -\k$,
reflecting an effectively mirror-symmetric coupling in the sense of condition (ii).

\subsection{Kuramoto model for large systems}
Numerical simulations show that $D_c$ depends weakly on system length $L$
for the two-dimensional Kuramoto model.
A heuristic argument suggests $D_c\sim 1/\ln L$~\cite{Pikovsky2003}, see SM text.

In the thermodynamic limit of infinite system size,
Kuramoto models with \textit{local}, short-range coupling
exhibit a conventional order-disorder transition as a function of noise strength
only in three- or higher-dimensional systems, but not in two-dimensional systems~\cite{Shinomoto1986,Sarkar2021a}.
This is essentially a consequence of the famous Mermin-Wagner theorem from statistical physics,
which rules out global order in two-dimensional systems with local coupling and continuous symmetries
\cite{Mermin1966}.
In fact, we can map the two-dimensional Kuramoto model with local sinusoidal coupling [Eq.~\eqref{eq:kuramoto}]
to the classical XY model of interacting spins in the plane with polar angles $\theta_j$
by switching to a co-rotating frame $\theta_j=\varphi_j - \omega_0 t$~\cite{Rouzaire2021,Sarkar2021a}.
Noise excites long-wavelength perturbation modes with long relaxation times
(so-called \textit{Goldstone modes}),
which rules out global synchronization in two-dimensional, infinitely extended systems.
These systems exhibit additional unusual features:
the spatial correlation function decays algebraically (with non-universal, parameter-dependent exponent);
moreover, topological defects can be easily excited in these systems
(and are in fact favored by entropy in the XY model).
Topological defects can form pairs below a characteristic noise strength,
corresponding to a Kosterlitz-Thouless transition, i.e., an unconventional phase transition.
In contrast, the Kuramoto model with \textit{global}, all-to-all coupling exhibits
a conventional second-order phase transition as a function of noise strength~\cite{Kuramoto1984,Strogatz2000,Son2010,Doerfler2014}.
Thus, there is a stark difference between Kuramoto models with global coupling
and two-dimensional models with local coupling,
previously studied for quenched frequency disorder~\cite{Sakaguchi1987,Strogatz1988,Lee2010, Ottino2016}
and noise \cite{Shinomoto1986, Sarkar2021a}.
Models of cilia carpets~\cite{Hong2020} ignoring this difference have limited predictive power.

Based on the correspondence between the cilia carpet model and the two-dimensional Kuramoto model
(as suggested by Fig.~\ref{figure3}
and demonstrated by the mapping from Eq.~\eqref{eq:kuramoto} to Eq.~\eqref{eq:kuramoto2}),
we expect that large carpets of coupled noisy cilia
likewise do not exhibit global synchronization, but instead local synchronized domains.

 \subsection{Discussion}

We investigated the role of active noise on metachronal synchronization in cilia carpets
using a multi-scale model.
Non-trivial synchronized states in the form of traveling waves persist even in the presence of noise,
featuring one dominant wave whose synchronized fraction exceeds those of all other possible waves.
Beyond a characteristic noise strength,
global synchronization is lost even in finite systems,
characterized by a rapid decline of spatial and temporal correlations.
Close to the characteristic noise strength,
we observe stochastic transitions between the synchronized and the disordered regime,
which one may regard as a high-dimensional analogue of phase-slips in pairs of noisy oscillators
\cite{Stratonovich1967,Goldstein2009}.

We highlighted similarities and differences of synchronization in cilia carpets
and the two-dimensional Kuramoto model of phase oscillators:
the cilia carpet model and the Kuramoto model
both exhibit a loss of global synchronization in finite systems beyond a characteristic noise strength.
As a small difference, the decline of spatial and temporal correlations
as a function of noise strength is more abrupt in the cilia carpet model.
Generally,
the temporal correlation function of a single oscillator features first a rapid decay,
which reflects small-amplitude fluctuations of this oscillator around a synchronized state,
followed by a slow decay, which reflects slow phase diffusion of a global phase associated to this synchronized state.
In addition, in the cilia carpet model,
we observe a peculiar oscillatory component of the temporal correlation function,
which reflects slow, noise-induced oscillations of the system's dynamics.
The period of these slow oscillations matches the imaginary part of the Lyapunov exponent
of the slowest-decaying perturbation mode of the dominant synchronized state, which becomes excited by noise.

The decline of global synchronization beyond a characteristic noise strength can be intuitively understood as follows:
in a limit of weak oscillator coupling and weak noise, one can linearize the dynamics around the dominant synchronized state.
This yields fundamental perturbation modes,
each of which has a characteristic relaxation time,
and whose dynamics is approximately described by a relaxation process driven by active noise
(Ornstein-Uhlenbeck process), see SM text for details.
Importantly, the relaxation time of the slowest-decaying perturbation mode diverges as $\sim L^2$
with system length $L$.
This can be analytically derived for the Kuramoto model with local coupling
and has been numerically shown for the cilia carpet model \cite{Solovev2020b}.
Thus, in larger systems perturbation modes with longer wave-length and longer relaxation times appear
\cite{Sakaguchi1987,Pikovsky2003}.
Assuming that a global synchronized state exists,
one can compute the fluctuations of an individual oscillator around this synchronized state
as a sum over all perturbation modes,
and thus arrive at a self-consistency condition
that the amplitude of these fluctuations is small in a globally synchronized state.
This heuristic argument suggests a condition for global synchronization in a two-dimensional system of system length $L$
in terms of the noise strength $D$ as $D<D_c(L)$
with a characteristic noise strength $D_c$ that depends weakly on system length $L$ as $D_c\sim 1/\ln L$.
As a consequence,  global synchronization is not possible in two-dimensional, infinitely-extended systems,
even for arbitrarily small noise strengths.
If a noise strength $D$ satisfies
$D_c(L_1)\ll D\ll D_c(L_2)$,
we may thus expect local synchronization on length scales $L_2$,
but not anymore on length scales $L_1$.
Indeed, previous simulations of the two-dimensional Kuramoto model
showed examples of such local synchronization \cite{Sarkar2021a}.

Despite the similarities,
there is one striking difference between the cilia carpet and the classical Kuramoto model:
in the cilia carpet, mirror-symmetry is broken:
This broken mirror-symmetry in the cilia carpet model manifests itself most clearly in a dominant wave with non-trivial wave vector
(which is rotated counter-clockwise relative to a reference direction
set by the direction of the effective stroke of the cilia beat).
In fact, the linear stability of different synchronization states with respective wave vectors $\k$
is symmetric under a reflection $\k\rightarrow -\k$ for the Kuramoto model
with mirror-symmetric coupling,
but highly asymmetric for the cilia carpet,
where Lyapunov exponents have the most negative real part close to the dominant wave.  The broken mirror-symmetry of the emergent dynamics in the cilia carpet model is a result of the chiral cilia beat pattern.

 Our analysis became possible by a multi-scale simulation approach \cite{Friedrich2012,Polotzek2013,Solovev2020a},
which describes a carpet of beating cilia as an array of phase oscillators,
with direction-dependent coupling functions that were calibrated
from detailed hydrodynamic simulations using a measured cilia beat pattern from \textit{Paramecium}
\cite{Machemer1972, Naitoh1984}.
Our approach allows to systematically vary the strength of active phase fluctuations,
whereas a previous study addressing synchronization in cilia carpets
employed a stochastic hydrodynamic simulation algorithm
and thus included noise only implicitly \cite{Elgeti2013}.

 For technical reasons, cilia spacing in our model ($a{=}18\,\micron$)
is larger than in real cilia carpets ($2\,\micron$ \cite{Machemer1972}),
similar to the dilute limit considered in most theoretical studies.
Therefore, we underestimate hydrodynamic interactions.
 Correspondingly, we consider lower noise strengths,
corresponding to quality factors $Q_c=\omega_0/(2D_c)$ that are approximately 10--30 times larger than previous measurements with quality factors in the range $Q=25$--$100$
\cite{Goldstein2011,Ma2014}.
  Real cilia carpets are characterized also by
quenched disorder of cilia position,
dispersity of intrinsic beat frequency,
and non-perfect alignment of cilia \cite{Guirao2010},
which should reduce the regularity of emergent metachronal waves,
in addition to active frequency jitter spotlighted here.
In the presence of these additional sources of disorder,
we expect qualitatively similar behavior,
albeit with lower characteristic noise strength (for the same cilia density).
In conclusion,
the dynamics of a cilia carpet in the presence of active noise
is similar to that of a two-dimensional Kuramoto model,  likewise displaying local but not global synchronization,
with the crucial difference that the dominant synchronization state is not the in-phase synchronized state,
but a so-called metachronal, traveling wave with non-trivial wave vector.

\subsection{Supplementary materials}

The manuscript is accompanied by online supplementary materials:
\textit{Supplementary material text} (SM text) and \textit{Supplementary movies}.

The SM text contains
details on numerical methods,
as well as auxiliary analytical and numerical results for the Kuramoto model
and the linearized Kuramoto model.

Two supplementary movies S1 and S2
show typical stochastic trajectories of the cilia carpet model
for a $48 \times 16$ lattice,
featuring synchronized and disordered dynamics, respectively.
Colored dots represent the deviations of cilia phases
$\Delta\varphi_j = \varphi_j + \k_{\mathrm{I}} \cdot \x_j - \langle \varphi_i + \k_{\mathrm{I}} \cdot \x_i \rangle$
from a perfect reference wave with wave vector $\k_{\mathrm{I}}$
at respective lattice positions $\x_j$,
where $\langle \cdot \rangle$ denotes the circular mean,
using the same color code as in Fig.~\ref{figure2}(b).
Noise strength for both movies was
$D = 0.17\,\mathrm{s}^{-1} \approx 0.94 D_c$
(same as Fig.~\ref{figure4},
with $D_c$ the characteristic noise strength for the $16 \times 16$ lattice).

\subsection{Acknowledgements}

AS and BMF are supported by the German National Science Foundation (DFG)
through
the \textit{Microswimmers} priority program (DFG grant FR3429/1-1 and FR3429/1-2 to BMF),
a Heisenberg grant (FR3429/4-1),
as well as
through the Excellence Initiative by the German Federal and State Governments
(Clusters of Excellence cfaed EXC-1056 and PoL EXC-2068).
We thank Christa Ringers and Nathalie Jurisch-Yaksi (NTNU), as well as all members of the `Biological Algorithms' group for stimulating discussions.

The data and the code accompanying this study are partially available online~\cite{gitall} or upon request.

\bibliography{cilia_carpet}

\clearpage

\appendix

\section{\large{Supplementary Material}}

{\noindent
Anton Solovev,
Benjamin M. Friedrich:\\
\textbf{Synchronization in cilia carpets and the Kuramoto model with local coupling:
breakup of global synchronization in the presence of noise}
}

\renewcommand{\theequation}{S\arabic{equation}}
\setcounter{equation}{0}
\renewcommand{\thefigure}{S\arabic{figure}}
\setcounter{figure}{0}
\renewcommand{\thetable}{S\arabic{table}}
\setcounter{table}{0}
\renewcommand{\thepage}{S\arabic{page}}
\setcounter{page}{1}

\section{Numerical methods}

Details on hydrodynamic simulations and the multi-scale modeling framework
are presented in the supplementary material of~\cite{Solovev2020b}.

\section{Numeric integration of stochastic equation of motion}

To numerically integrate the stochastic equations of motion,
Eq.~(\ref{eq:noise}),
we used an Euler-Maruyama scheme
with fixed time step $\Delta t = T_0 / 100$
(where $T_0=2\pi/\omega_0 = 31.25\,\mathrm{ms}$ represents the intrinsic beat period of a single cilium).
In each time-step, we invert the matrix $\Gmat$ of generalized hydrodynamic friction coefficients using sparse matrix algorithms.

\subsection{Ensemble-averages in stochastic simulations}
For each noise strength $D$,
we simulated $n=200 - 400$ trajectories, initialized from uniformly distributed random initial conditions.

\begin{figure}
\includegraphics[width=\linewidth]{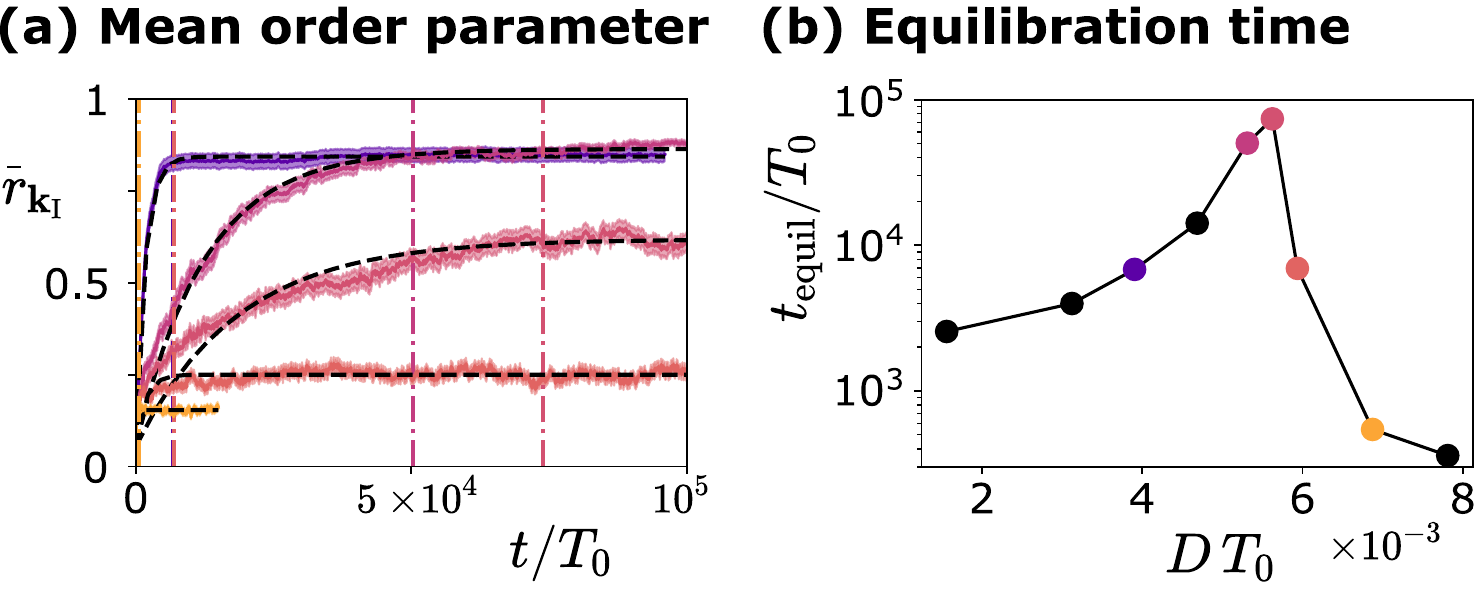}
\caption{
\textbf{Equilibration time of stochastic simulations.}
(a)
Mean order parameter $\overline{r}_{\k_{\mathrm{I}}}$ as a function of time $t$
for different noise strengths $D$
(purple: $D=0.125\,\mathrm{s}^{-1}$, shades of red: $D=0.17,\,0.18,\,0.19\,\mathrm{s}^{-1}$, respectively,
orange: $D=0.22\,\mathrm{s}^{-1}$; see also panel (b) for colors used).
Vertical lines indicate the computed equilibration time $t_\mathrm{equil}$ for each case.
Black dashed lines show the fit $r_{\mathrm{fit}}(t)$
according to Eq.~\eqref{eq:SI_equil_time}.
(b)
The equilibration time $t_\mathrm{equil}$ depends sensitively on noise strength $D$,
and in particular increases close to the characteristic noise strength $D_c$.
}
\label{figureS11}
\end{figure}

To calculate ensemble-averaged quantities at steady state (e.g., synchronized fraction, correlations $C(\Delta t)$, $S(\d)$),
we averaged over this ensemble of trajectories after an equilibration time $t_\text{equil}$.
To determine
$t_\text{equil}=t_\text{equil}(D)$
for each noise strength $D$,
we first computed the instantaneous mean order parameter of the ensemble
$\overline{r}_{\vec{k}_{\mathrm{I}}}(t) = \langle r_{\vec{k}_{\mathrm{I}}}(t)\rangle$.
The time course of this mean order parameter was well-approximated by a two-parameter fit
\begin{equation}
r_\text{fit}(t; \tau, r_\infty) = r_0 +  (r_{\infty} - r_0)(1 - e^{- t/ \tau})\quad,
\label{eq:SI_equil_time}
\end{equation}
with steady-state value $r_\infty$, relaxation time-scale $\tau$, and
initial value $r_0 = \overline{r}_{\ki}(0)$,
corresponding to the mean order parameter for uniformly distributed phase vectors,
see Fig.~\ref{figureS11}(a).
Note that
$r_0 \sim 1 / \sqrt{N}$ for all $\k$.
We used the heuristic $t_{\text{equil}} = 4 \tau$,
which corresponds to $|r_\infty - \overline{r}_\vec{k}(t) | / (r_\infty - r_0) < e^{-4}  \approx 0.02$.
We found that $t_\text{equil}$ depends sensitively on noise strength $D$,
taking values in the range of $10^2 - 10^5\,T_0$,
see Fig.~\ref{figureS11}(b).
Additionally, we observed that this $t_{\mathrm{equil}}$ grows with the number of oscillators $N$,
which makes it increasingly difficult to study larger systems.

\subsection{Synchronized fraction}
We computed the synchronized fraction near a traveling wave solution $\FP_\k$
with wave vector $\k$
as the fraction of time that simulated trajectories spent in the vicinity of the corresponding wave, quantified as $r_\k(\F) > r^*$.
Here, we used the threshold $r^* = \sqrt{2}/2$.
This choice of threshold is motivated as follows:
the order parameters $r_\k$ are
closely related to Fourier transform on the lattice;
therefore, $\sum_{k \in \K} r_\k(\F)^2 = 1$
as a consequence of Plancherel's theorem.
Hence, $r_\k  > \sqrt{2}/2$ guarantees $ r_{\l} < \sqrt{2}/2$ for all $\l \in \K / \{ \k\}$,
i.e., the choice of threshold ensures that the neighborhoods are mutually disjoint.
Note that these neighborhoods comprise only a tiny ($<10^{-10}$) fraction of phase space volume.

\subsection{Simulations on a larger lattice}
For Fig.~\ref{figure4}, we considered a triangular lattice with dimensions $N_x = 48$ and $N_y = 16$,
while other simulation parameters were identical to the case of the  $16 \times 16$ lattice studied in Fig.~\ref{figure3}.
For $D=0.17\,\mathrm{s}^{-1}$,
we integrated $n=32$ trajectories for a time $t_{\max}=4 \cdot 10^{5}\, T_0$.
To speed up equilibration,
instead of drawing uniformly distributed random initial conditions,
we used the last time points of simulated trajectories for the $16\times 16$ lattice for the same noise strength
to define initial conditions for the $48 \times 16 $ lattice.
As a result, we could not use the equilibration time formula
Eq.~\eqref{eq:SI_equil_time},
as the mean order parameter $\overline{r}_{\ki}(t)$ is now decreasing,
instead of increasing as in the case of uniformly distributed random initial conditions.
Instead, we visually confirmed that after $t > t_{\max} /2 $,
the ensemble-averaged order parameter $\overline{r}_{\ki}(t)$ fluctuates, but does not drift,
and used $t_\text{equil} = t_{\max} /2$ as the equilibration time.

\section{Comparison with Kuramoto model on a triangular lattice}

\subsection{Choice of coupling strength $K$}
To compare the cilia carpet model with the Kuramoto model
on the $16 \times 16$ triangular lattice
(Fig.~\ref{figure3} in the main text),
we have chosen the coupling strength $K$ as follows.
We define a characteristic noise strength $D^\text{Kuramoto}_c$
as the noise strength at which
the sum of synchronized fractions $R_\k$ for all waves $\k$ equals $50\%$.
Using an arbitrary, but fixed value of $K$,
we found
$D^\text{Kuramoto}_c \approx 0.25\, K$.
We require that the characteristic noise strength in the Kuramoto model should match
the characteristic noise strength for the cilia carpet model
$D^\text{cilia}_c \approx 0.18\,\mathrm{s}^{-1}$;
rescaling $K$ yields the parameter choice
\begin{equation}
K
= 4\, D^{\text{cilia}}_c
= 0.72\, \mathrm{s}^{-1}
\period
\end{equation}

\subsection{Characteristic noise strength as a function of the system size}
\begin{figure}
\begin{center}
\includegraphics[width=\linewidth]{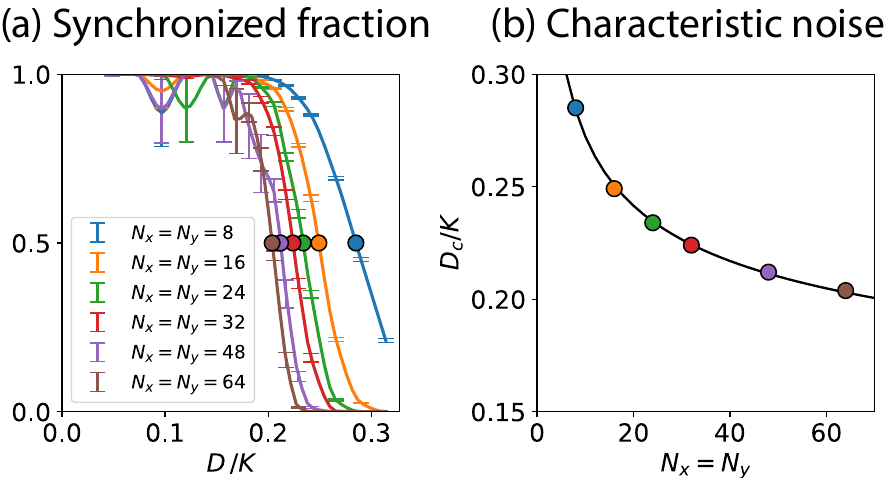}
\end{center}
\caption[]{
(a) Synchronized fraction
    for in-phase synchronization
    in the Kuramoto model on two-dimensional triangular lattice of different size (solid lines).
    Characteristic noise strength (colored dots) was defined
    as the noise strength $D$, where the synchronized fraction of the dominant synchronized state ($\k = \zero$) equals $50\%$.
(b) Characteristic noise strength $D_c$ as a function of system size $N_x$ (colored dots).
    Solid line shows a fitted function $f(D; a, b) = a ( \log(D) + b)^{-1}$
}
\label{figureS12}
\end{figure}

We computed synchronized fractions
for the synchronized regime near the in-phase periodic solution ($\k = \zero$)
for different noise strengths and various system lengths $L /a = N_x = N_y$.
Fig.~\ref{figureS12}(a) shows
the synchronized fraction as a function of the dimensionless noise strength $D/K$.
For each system length,
we defined a characteristic noise strength $D_c$
as the noise strength where the curve of synchronized fraction intersects $1/2$.
Thus, we obtain $D_c$ as function of system length $L$, see Fig.~\ref{figureS12}(b).
We observe that $D_c$ slowly decreases with the system length,
in agreement with the theoretical scaling $D_c \sim (\ln L)^{-1}$
obtained for the linearized system,
see SM section `Continuum limit'.

\subsection{Kuramoto model with nearest-neighbor sinusoidal coupling in $d$ dimensions in the presence of noise}

One can consider a Kuramoto model of noisy phase oscillators with identical frequencies
on a hyper-cubic lattice in $d$-dimensional space (with lattice spacing $a$ and lattice positions $\x_i$).
We assume local nearest-neighbor coupling, i.e., each oscillator with phase variable $\varphi_i$ is coupled to its $m=2d$ nearest neighbors (enumerated by an index set $\N_i$) with total coupling strength $K$,
\begin{equation}
\label{eq:Kuramoto_sine}
\dot{\varphi}_i = \omega_0 - \frac{K}{m} \sum_{j\in\N_i} \sin(\varphi_i-\varphi_j) + \xi_i \quad.
\end{equation}
Here, $\xi_i(t)$ denotes uncorrelated Gaussian white noise with
$\langle \xi_i(t)\xi_j(t')\rangle = 2D\,\delta_{ij}\,\delta(t-t')$.
We impose periodic boundary conditions.

In the absence of noise,
linear stability analysis yields a set of fundamental perturbation modes~\cite{Solovev2020b}
\begin{equation}
\label{eq:delta_l}
\boldsymbol{\Delta}_\l : \delta_j = \varepsilon \exp( -  i\, \l\cdot\x_j) \text{ for  } \l\in\mathbbm{K}\setminus\{\boldsymbol{0}\}
\comma
\end{equation}
the corresponding dimensionless Lyapunov exponents read
\begin{equation}
\label{eq:lambda_l}
\lambda_\l = -\frac{K T_0}{m} \sum_{j\in\N_i} \cos(\k\cdot\x_{ij}) \left[ 1 - \cos(\l\cdot\x_{ij}) \right] \quad,
\end{equation}
where we used short-hand $\x_{ij} = \x_j-\x_i$.

In the limit of weak noise, $D\ll K$,
phases fluctuate around $\ol{\varphi}$ and
we can linearize Eq.~\eqref{eq:Kuramoto_sine}
\begin{equation}
\label{eq:Kuramoto_linear}
\dot{\varphi}_i = \omega_0 - \frac{K}{m} \sum_{j\in\N_i} \left(\varphi_i-\varphi_j\right) + \xi_i \quad.
\end{equation}
We introduce the discrete set of Fourier modes
\begin{equation}
\wt{\varphi}_\k = a^d\, \sum_j \varphi_j(t) \exp(-i\,\x_j\cdot\k)
\quad,
\end{equation}
where $k\in\K$ is an element of the first Brillouin zone of the reciprocal lattice.
We introduce short-hand $\K^\ast = \K\setminus\{\zero\}$,
noting that the wave modes $\k=\zero$ will behave different.
The inverse Fourier transform reads
\begin{equation}
\varphi_j(t) = V^{-1} \sum_{\k\in\mathbbm{K}} \wt{\varphi}_\k(t) \exp(i\, \x_j\cdot\k) \quad,
\end{equation}
where $V=a^d \,N$ is the $d$-dimensional volume of the system
and $N=\Pi_{i=1}^d N_i$ the number of oscillators
with system size $N_1 \times \ldots \times N_d$.
By conjugation, $\wt{\varphi}^\dagger_\k=\wt{\varphi}_{-\k}$.
The Fourier coefficients are closely related to the Kuramoto order parameters $r_\k$ introduced in Eq.~\eqref{eq:rk}
as $r_\k \approx V^{-1} |\wt{\varphi}_{-\k}| = V^{-1} |\wt{\varphi}_\k|$,
provided $D\ll K$, hence $|\varphi_j|\ll 1$.

\subsection{Dynamics of Fourier modes}
The Fourier-transform of the linearized dynamics, Eq.~\eqref{eq:Kuramoto_linear},
shows that each Fourier mode $\wt{\varphi}_\k$ fluctuates as an independent complex Ornstein-Uhlenbeck process around zero
\begin{equation}
\frac{d}{dt} \wt{\varphi}_\k = - \frac{1}{\tau_\k}\, \wt{\varphi}_\k + \xi_\k(t) \text{ for } \k\in\K^\ast\quad,
\end{equation}
where
$\xi_\k(t)$ is isotropic complex Gaussian white noise with
$\langle \xi_\k(t)\xi^\ast_\l(t+\Delta t)\rangle = 2\sigma^2 V\,\delta_{\k,\l}\,\delta(t-t')$,
where $\sigma^2=D a^d$.
Note $\langle \xi_\k(t)\xi_\k(t+\Delta t)\rangle = 0$, because contributions from the real and imaginary part of $\wt{\varphi}_\k$ exactly cancel.
The relaxation times $\tau_\k$ read
\begin{equation}
\tau_\k = \frac{m}{2K} \left( d - \sum_{n=1}^d \cos( a k_n ) \right)^{-1} \quad.
\label{eq:tau_k}
\end{equation}
The derivation of $\tau_\k$ uses Eq.~\eqref{eq:Kuramoto_linear} and the shift theorem of the discrete Fourier transform.
The cross-correlation of the Fourier coefficients is thus given by
\begin{align}
\langle \wt{\varphi}_\k(t) \wt{\varphi}^\ast_\l(t+\Delta t) \rangle
&=
\var(\wt{\varphi}_\k)\,\delta_{\k,\l}\, \exp(-|\Delta t|/\tau_\k),
\end{align}
with variance
\begin{equation}
\var(\wt{\varphi}_\k) = \sigma^2 V\,\tau_\k \quad.
\end{equation}
The Fourier mode with $\k=\zero$ represents an exception,
and is described by a diffusion process with drift
\begin{equation}
\frac{d}{dt} \wt{\varphi}_\zero = \omega_0 V + \xi_\zero(t) \quad,
\end{equation}
where
$\xi_\zero(t)$ is Gaussian white noise with
$\langle \xi_\zero(t)\xi_\zero(t+\Delta t)\rangle = 2\sigma^2 V\,\delta(t-t')$.
This zeroth Fourier mode is closely related to the \textit{global phase}
$\ol{\varphi}=\sum_j \varphi_j / N$ as
\begin{equation}
\ol{\varphi} = V^{-1} \,\wt{\varphi}_\zero \quad.
\end{equation}
The phase correlation function of the global phase reads \cite{Ma2014}
\begin{multline}
C_{\ol{\varphi}}(\Delta t)
= |\langle \exp i\, [\ol{\varphi}(t+\Delta t) - \ol{\varphi}(t)] \rangle | \\
= \exp\left( -\frac{D}{N}|\Delta t|\right) \quad,
\end{multline}
i.e., $\ol{\varphi}$ exhibits effective phase diffusion with effective diffusion coefficient $D/N$
(in addition to its deterministic drift with $\langle \ol{\varphi}(t+\Delta t)-\ol{\varphi}(t) \rangle = \omega_0 \Delta t$).

We are interested in the deviations
$\delta_j=\varphi_j - \ol{\varphi}$
of the phases of the individual oscillators from the global phase.
For the autocorrelation function of these deviations, we find
\begin{align}
\langle \delta_j(t) \delta_j(t+\Delta t) \rangle
&=
V^{-2}\!\!\! \sum_{\k\in\K^\ast} \!\!\! \var(\wt{\varphi}_\k)\, \exp(-|\Delta t|/\tau_\k) \quad,
\end{align}
with relaxation times $\tau_\k$ given in Eq.~(\ref{eq:tau_k}).

\subsection{Temporal correlations}
Similarly, we can compute the phase correlation function of $\delta_j$.
We first note the phase correlation function (also called moment-generating function, characteristic function, circular autocorrelation function)
of an Ornstein-Uhlenbeck process $\delta(t)$ given by
$\dot{\delta} = - \delta/\tau + \xi$,
$\langle \xi(t)\xi(t') \rangle = 2D\,\delta(t-t')$,
with relaxation time $\tau$ and variance $D\tau$ as%
\footnote{
For a proof, note that we write $x(t+\Delta t)-x(t)=A+B$ as a sum of two independent random variables,
$A=x(t) [ \exp(-\Delta t/\tau) - 1]$,
and $B=\int_t^{t+\Delta t}\! dt'\, \xi(t')\,\exp[- (t+\Delta t - t')/\tau]$,
where $A$ and $B$ are normal distributed random variables with zero mean and respective variances,
$D\tau\, [1-\exp(-\Delta t/\tau)]^2$ and $D\tau\,[1-\exp(-2\Delta t/\tau)]$.
Since $\langle \exp i\, \Xi \rangle = \exp[-\langle \Xi^2 \rangle/2 ]$
for normally distributed random variables with zero mean,
we conclude
\begin{multline*}
\langle \exp i\, [ x(t+\Delta t) - x(t) ] \rangle
= \langle \exp i A \rangle \cdot \langle \exp i B \rangle \\
= e^{- D\tau\, [1-\exp(-\Delta t/\tau)]^2/2} \cdot e^{- D\tau [1-\exp(-2\Delta t/\tau)]/2}\quad,
\end{multline*}
from which the assertion follows.
}
\begin{multline}
C_\delta(\Delta t)
= |\langle \exp i\, [\delta(t+\Delta t) - \delta(t)]  \rangle| = \\
\exp\left( - D\tau \left[ 1- \exp\left(-\frac{|\Delta t|}{\tau}\right) \right] \right) \quad.
\end{multline}
This phase correlation function converges to a non-zero limit value
\begin{equation}
\lim_{\Delta t \to \infty} C_\delta(\Delta t)
= \exp{\left( - D \tau \right)}
\quad.
\end{equation}
We now compute the phase correlation function of $\delta_j$.
Without loss of generality, it suffices to consider
$\delta_0$ at the origin $\r_0=\zero$ due to the translation symmetry of the oscillator lattice.
Since $\delta_0(t)$ is a sum of independent complex Ornstein-Uhlenbeck processes,
namely the $\wt{\varphi}_\k(t)$,
it follows that the circular autocorrelation function of $\delta_0$
is a product of their respective circular autocorrelation functions
\begin{multline}
C'(\Delta t) =
|\langle \exp i\, [\delta_0(t+\Delta t) - \delta_0(t)]  \rangle| = \\
\prod_{\k\in\K^\ast}
  \exp\left(
    - \frac{D}{N}\,\tau_\k     \left[
      1- \exp\left(-\frac{|\Delta t|}{\tau_\k}\right)
    \right]
  \right) \quad.
\end{multline}
Because global phase $\ol{\varphi}$ and phase deviation $\delta_0$ are independent random variables,
the phase correlation function
$C(\Delta t)=|\langle \exp i\,[\varphi_j(t+\Delta t) - \varphi_j(t)] \rangle |$
of
$\varphi_0 = \ol{\varphi} + \delta_0$
is given as the product of $C_{\ol{\varphi}}(\Delta t)$ and $C_\delta(\Delta t)$.
We thus understand the behavior of temporal correlations on short and long time-scales:
\begin{itemize}
\item \textit{Short-time dynamics.}
For short times, $\Delta t \lesssim \tau_\k$,
$C(0,\Delta t)$ will rapidly decay to a limit value
$C_\infty = \Pi_{\k\in\Kp} \exp(-D\tau_\k) = \exp[-\var(\delta_j)]$
due to fluctuations of individual phases around the global phase $\ol{\varphi}$,
where
$\var(\delta_j)=V^{-2} \sum_{\k\in\Kp} \var(\wt{\varphi}_\k)$
is the variance of $\delta_j$.
This decay is a superposition of exponentials with a spectrum of relaxation time-scales given
by $\tau_\k$ for $\k\in\Kp$.
\item
\textit{Long-time dynamics.}
For long times, $\Delta t \gg \tau_\k$, $C(0,\Delta t)$ decays as
$C(0,\Delta t) \approx R_\infty \, \exp[- (D/N) \Delta t ]$,
reflecting diffusion of the global phase $\ol{\varphi}$.
\end{itemize}
In summary,
the temporal correlation function of the phase $\varphi_j$
is characterized by an initial fast decay due to fluctuations of individual phases around the global phase,
and a subsequent slow decay due to phase diffusion of the global phase itself
\begin{equation}
\boxed{
C(\Delta t) \approx^{\strut}
\underbrace{
C'(\Delta t)
}_\text{fast}
\,
\underbrace{
\exp\left(-D|\Delta t|/N\right)
}_\text{slow}
}
\quad,
\end{equation}
where
$C'(0)=1$, and
$C'(\Delta t)\rightarrow C_\infty = \exp[-\var(\delta_0)]$ for $\Delta t\gg \tau=\max \tau_\k$.

\subsection{Spatial correlation}
\begin{figure}
\includegraphics[width=0.8\linewidth]{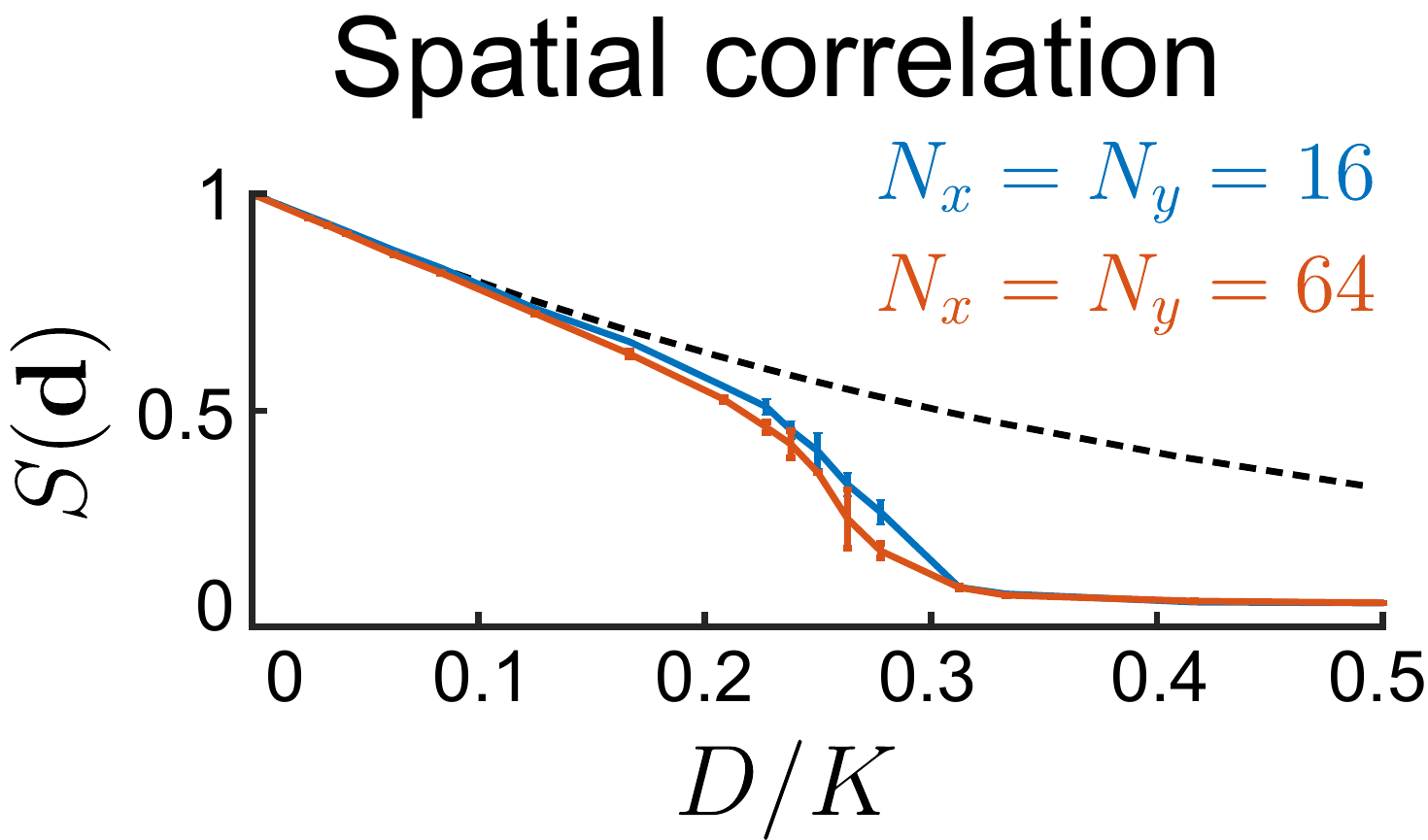}
\caption{
\textbf{Spatial correlation} $S(\d)$
as a function of normalized noise strength $D/K$
for displacement vector $\d=7a\,\e_x$
and
two system sizes: $N_x=N_y=16$ (blue) and $N_x=N_y=64$ (golden),
for oscillators on a rectangular lattice.
Dashed line shows analytical result from Eq.~(\ref{eq:Sd_ana}), using $N_x=N_y=64$.
Average over $n=2$ trajectories initialized at $\varphi_j=0$ for all $j$,
equilibrated for $t_\mathrm{equil}=10^2/D$, time step $dt=10^{-4}/D$, total integration time $5\cdot 10^2/D$}
\label{figureS13}
\end{figure}

We compute the spatial correlation function
$S(\d) = \langle \exp i\, [\varphi(\x_j + \d,t) - \varphi(\x_j,t) ] \rangle$.
By translational symmetry,
$S(\d)$ is independent of $j$ and we may choose $j=0$ with $\r_0 = \zero$ without loss of generality.
The inverse Fourier transform thus gives
$ \varphi(\x_j + \d,t) - \varphi(\x_j,t) = V^{-1}\sum_{\k\in\K} \wt{\varphi}_\k [ \exp( i\, \d\cdot\k ) - 1]$
by the shift theorem of the discrete Fourier transform.
Different Fourier modes $\wt{\varphi}_\k$ and $\wt{\varphi}_\l$ are independent,
as are their real and imaginary parts,
except for $\k$ and $-\k$,
which form a conjugate pair, $\wt{\varphi}_\k^\dagger = \wt{\varphi}_{-\k}$.
Hence%
\footnote{
We write
$\mathcal{F} = V^{-1} \left( \wt{\varphi}_\k\, [ e^{i\,\d\cdot\k} - 1 ] + \wt{\varphi}_{-\k}\, [ e^{-i\,\d\cdot\k} - 1 ] \right)$
as
\begin{equation}
\mathcal{F}
= 2V^{-1} \left(
\wt{\varphi}'_\k\, [ \cos(\d\cdot\k) - 1 ]
-\wt{\varphi}''_\k\, \sin(\d\cdot\k)
\right)\quad,
\end{equation}
where
$\wt{\varphi}_\k=\wt{\varphi}'_\k + i\wt{\varphi}''_\k$ is the decomposition into real and imaginary part.
Thus, the variance of $\mathcal{F}$ reads
\begin{align}
&\var\left( \mathcal{F} \right)
\notag \\
&= 4V^{-2} \left(
\var(\wt{\varphi}'_\k)\, [ \cos(\d\cdot\k) - 1 ]^2
+\var(\wt{\varphi}''_\k)\, \sin^2(\d\cdot\k)
\right)
\notag \\
&= 4V^{-2} \left(
\frac{\sigma^2 V\tau_\k}{2}\, [ \cos(\d\cdot\k) - 1 ]^2
+\frac{\sigma^2 V\tau_\k}{2}\, \sin^2(\d\cdot\k)
\right)\quad,
\notag \\
&= \frac{4D\tau_\k}{N} [ 1 - \cos(\d\cdot\k) ] \quad.
\end{align}
Moreover, $\mathcal{F}$ is a normal distributed random variable with $\langle\mathcal{F}\rangle=0$, hence
\begin{equation}
\left\langle \exp i \mathcal{F} \right\rangle = \exp[ - \var(\mathcal{F})/2 ]
\quad.
\end{equation}
}
\begin{align*}
S(\d)
& = \left|\left\langle
e^{ i \,[
\varphi(\x_j+\d,t) - \varphi(\x_j,t)
] }
\right\rangle\right| \\
&=
\prod_{\pm\k\in \Kp} \!\!\!\!\!{}'\,
\left\langle
\exp i\left(
V^{-1}\, \wt{\varphi}_\k\, [ e^{i\,\d\cdot\k} - 1 ] + \mathrm{c.c.}
\right)
\right\rangle \\
&=
\prod_{\pm\k\in \Kp} \!\!\!\!\!{}'\,
\exp\left(
- \frac{2D}{N}\,\tau_\k
\left[
1- \cos(\mathbf{d}\cdot\mathbf{k})
\right]
\right) \quad.
\end{align*}
Here, $\Pi'$ denotes a product over pairs of conjugate wave vectors, $\k$ and $-\k$.
We conclude for the spatial correlation function
\begin{equation}
\boxed{
S(\d) =
\prod_{\k\in \Kp}
\exp
\left(
- \frac{D}{N}\,\tau_\k
\left[
1- \cos(\mathbf{d}\cdot\mathbf{k})
\right]
\right)
}
\quad.
\label{eq:Sd_ana}
\end{equation}
This analytical result for $S(\mathbf{d})$
agrees with simulation results for the Kuramoto model with local sinusoidal coupling,
Eq.~(\ref{eq:Kuramoto_sine}), in the limit of weak noise, $D\ll K$, but deviates for stronger noise, see Fig.~\ref{figureS13}.
Note that the spatial correlation function $S(\d)$ is largely independent of system size for fixed separation vector $\d$,
as expected,
whereas the mean order parameter $\ol{r}_\zero$ decreases with system size.

\section{Continuum limit\\ (cf.~Pikovsky et al., \cite[Chapter 11]{Pikovsky2003})}

Following \cite{Pikovsky2003},
we consider an oscillator lattice with weak isotropic nearest-neighbor coupling,
and characterize noise-induced fluctuations in a continuum limit.
Specifically,
we consider a $d$-dimensional lattice $\Lambda\subset\mathbbm{R}^d$ with lattice spacing $a$,
where identical noisy phase oscillators with phase variables $\varphi_i$,
intrinsic frequency $\omega_0$ and noise strength $D$
are located at nodes $\x_i\in\Lambda$.
Each oscillator is coupled to its $2d$ nearest neighbors (enumerated by an index set $\N_i$),
with a generic $2\pi$-periodic coupling function $c$ and small coupling strength $\varepsilon>0$
\begin{equation}
\label{eq:oscillator_lattice}
\dot{\varphi}_i = \omega_0 + \varepsilon \sum_{j\in\N_i} c(\varphi_i-\varphi_j) + \xi_i \quad.
\end{equation}
Here, $\xi_i(t)$ denotes uncorrelated Gaussian white noise with
$\langle \xi_i(t)\xi_j(t')\rangle = 2D\,\delta_{ij}\,\delta(t-t')$.
We perform a continuum limit $a\rightarrow 0$
with corresponding renormalization of coupling and noise strength, $\varepsilon$ and $D$, respectively,
such that $\gamma = \varepsilon a^2$ and $\sigma^2=D a^d$ remain constant.
We obtain a partial differential equation for the phase field $\varphi(\x,t)$,
valid for long-wave perturbations of long-wave solutions%
\footnote{
For a rigorous derivation of Eq.~\eqref{eq:continuum_limit},
one should restrict the analysis to phase fields that vary only on long length scales $d\gg a$,
and subsequently perform a Taylor expansion to second order in the small dimensionless parameter $a/d$.
In the theory of stochastic partial equations, Eq.~\eqref{eq:continuum_limit} is known as a stochastic heat equation (for $\beta=0$),
whose solutions are (almost) $\alpha$-H{\"o}lder continuous, hence the differential operator $\nabla^2$ is well defined.
}
\begin{equation}
\label{eq:continuum_limit}
\frac{\partial}{\partial t} \varphi(\x,t) = \omega_0 + \alpha \nabla^2 \varphi(\x,t)
+ \beta \left[ \nabla \varphi(\x,t) \right]^2 + \xi(\x,t) \quad.
\end{equation}
Here, $\xi(\x,t)$ denotes uncorrelated Gaussian white noise%
\footnote{
Discretization of the stochastic partial differential equation Eq.~\eqref{eq:continuum_limit}
should yield again the system of coupled stochastic differential equations Eq.~\eqref{eq:oscillator_lattice}
with appropriately rescaled coupling and noise strengths.
To show this,
we introduce the discrete set of variables
$\varphi_i(t) = a^{-d} \int_{V_i} \!d\x\, \varphi(\x,t)$
that average the continuous field $\varphi(\x,t)$
over a $d$-cube unit cell $V_i$ centered at the lattice point $\x_i$.
Indeed, the stochastic part
$\Xi_i=\int_t^{t+\Delta t} dt' \int_{V_i} \!d\x\, \xi(\x,t)$
of the increment
$\varphi_i(t+\Delta t)-\varphi_i(t)$
is a normally distributed random variable with
zero mean, $\langle \Xi_i\rangle = 0$, and variance
$\langle \Xi_i\Xi_j \rangle =
a^{-2}\,2\sigma^2\,\Delta t\,a^d = 2D\,\Delta t$,
which equals $\int_t^{t+\Delta t} \!dt'\, \xi_i(t)$.
}
with
$\langle \xi(\x,t)\xi(\x',t')\rangle = 2\sigma^2\,\delta(\x-\x')\delta(t-t')$,
and the differential operators are to be interpreted in a suitably smoothed sense.
The effective parameters $\alpha$ and $\beta$ are given by
$\alpha = \gamma\, c'(0)$ and $\beta = \gamma\, c''(0)$.
While $\alpha$ represents an effective coupling strength,
the parameter $\beta$ characterizes a dispersion relation of plane wave solutions of Eq.~\eqref{eq:continuum_limit}
in the noise-free case with $D=0$ as
\begin{equation}
\label{eq:continuum_limit_dispersion}
\varphi(\x,t) = \omega_\k\, t - \k\cdot\x \quad, \quad \omega = \omega_0 + \beta\, |\k|^2 \quad.
\end{equation}
Eq.~\eqref{eq:continuum_limit} is also known as the Kardari-Parisi-Zhang equation
in the theory of roughening interfaces;
its linearization with $\beta=0$ yields the Edwards-Wilkinson equation,
which represents a special case of the complex Ginzburg-Landau equation for the case of constant oscillator amplitude.
We now restrict our analysis to a finite system of system length $L$
represented by a $d$-cube of volume $V=L^d$, and impose periodic boundary conditions.
We can then Fourier-transform the linearization of Eq.~\eqref{eq:continuum_limit} with $\beta=0$,
and obtain uncoupled Langevin equations for each Fourier mode
$\wt{\varphi}_\k = \int_V \! d\x\, \varphi(\x,t) \exp(-i\x\cdot\k)$
\begin{equation}
\label{eq:continuum_limit_fourier}
\frac{d}{dt} \wt{\varphi}_\k
= \omega_0 V\, \delta_{\k,\boldsymbol{0}} - \alpha |\k|^2\,\wt{\varphi}_\k + \xi_\k(t) \quad.
\end{equation}
Here, $\xi_\k(t)$ denote complex-valued uncorrelated Gaussian white noise with
$\langle \xi_\k(t)\xi_{\k'}^\ast(t') \rangle = 2\sigma^2 V\,\delta_{\k,\k'}\,\delta(t-t')$,
where the star denotes the complex conjugate.
Note that Eq.~\eqref{eq:continuum_limit_fourier} could have been equivalently derived as the
Fourier transform of the continuum limit of the linearized Kuramoto model, Eq.~\eqref{eq:Kuramoto_linear}.
Hence, each Fourier mode $\wt{\varphi}_\k$ fluctuates as a complex Ornstein-Uhlenbeck process with variance
$
\mathrm{var}(\wt{\varphi}_\k)
= \langle |\wt{\varphi}_\k|^2 \rangle - |\langle \wt{\varphi}_\k \rangle|^2
= \sigma^2\, V\, \tau_\k
$
around its mean $\omega_0 V\,\delta_{\k,\mathbf{0}}$
with relaxation time $\tau_\k = (\alpha |\k|^2)^{-1}$.
Introducing the long-wavelength (infrared) cutoff
$k_\mathrm{IR} = 2\pi/L$, where $L$ denotes system length, and
the short-wavelength (ultraviolet) cutoff
$k_\mathrm{UV} = \pi/a$, where $a$ denote the lattice spacing,
we find for the total variance of the phase field in real space
(the surface area of the $d$-dimensional unit ball is denoted $A_d$
with $A_1=2$, $A_2=2\pi$, $A_3=4\pi$, \ldots)%
\footnote{
The inverse Fourier transform reads
$\varphi_j(t) = V^{-1} \sum_{\k\in\mathbbm{K}} \wt{\varphi}_\k(t) \exp(i\, \x_j\cdot\k)$,
where we set
$\varphi(\x,t) = a^d\sum_j \varphi(\x_j,t)\,\delta(\x-\x_j)$
and used $V=Na^d$.
}
\begin{align}
\langle [ \varphi(\x,t) - \ol{\varphi}(t) ]^2 \rangle_t
&=
\frac{1}{V^2} \sum_{\k\in\mathbbm{K}\setminus\{\boldsymbol{0}\}} \mathrm{var}(\wt{\varphi}_\k) \notag \\
&\approx
\frac{1}{V(2\pi)^d}
\int_{|\k|=k_\mathrm{IR}}^{|\k|=k_\mathrm{UV}} \!d\k\, \mathrm{var}(\wt{\varphi}_\k) \notag \\
&=
\frac{\sigma^2 A_d}{\alpha(2\pi)^d}\,\int_{k_\mathrm{IR}}^{k_\mathrm{UV}} dk\, k^{d-1} k^{-2} \notag \\
&\approx
\begin{cases}
\frac{1}{2\pi^2} \frac{D a}{\alpha} L
& d=1 \\[1mm]
\frac{1}{2\pi} \, \frac{D a^2}{\alpha} \ln\left( \frac{L}{2a} \right) & d=2 \\
\mathrm{constant} & d\ge 3 \quad.
\end{cases}
\end{align}
Here,
$\ol{\varphi}(t)
= \wt{\varphi}_\mathbf{0}(t) / V
= \int \!d\x\,\varphi(\x,t) / V$
can be interpreted as global phase of the oscillator array.
In particular, for $d\le 2$,
the fluctuation amplitude diverges for $L\rightarrow\infty$.
For $d=2$,
we expect $D_{c} \sim 1 / \ln L$
by equating the fluctuation amplitude to a threshold value.

As a side-note, if we consider the spatial Fourier transform
$\wt{\varphi}_\k=a^d \sum_j \varphi_i \exp(-i \x_j\cdot\k )$
for a discrete lattice of oscillators with phases $\varphi_j$ at positions $\x_j$
instead of a continuous field $\varphi(\x,t)$,
then Eq.~\eqref{eq:continuum_limit_fourier} slightly changes for short wavelengths:
for a square lattice in $d$ dimensions with lattice spacing $a$, we have
$d\wt{\varphi}/dt = \omega_0 V\delta_{\k,\boldsymbol{0}} - \wt{\varphi}_\k / \tau_\k + \xi_\k$,
where the relaxation time now reads
$\tau_\k = (\alpha\, 2[ d - \sum_{l=1}^d \cos( a k_l )]/a^2 )^{-1}$.
For $|\k|\ll 1/a$, this formula agrees of course with the continuum limit Eq.~\eqref{eq:continuum_limit_fourier}.

\end{document}